
%
%

\def\preprint{Y}        
\input jnl
\input figuredefs
\input reforder
\if \preprint Y \twelvepoint\oneandathirdspace \fi
\title Simple Hadronic Matrix Elements with Wilson Valence
Quarks and Dynamical Staggered Fermions
at ${\bf 6/g^2=5.6}$

\author
Khalil M.~Bitar,${}^{(1)}$ T.~DeGrand,${}^{(2)}$ R.~Edwards,${}^{(1)}$
Steven Gottlieb,${}^{(3)}$ U.~M.~Heller,${}^{(1)}$ A.~D.~Kennedy,${}^{(1)}$
J.~B.~Kogut,${}^{(4)}$ A.~Krasnitz,${}^{(3)}$ W.~Liu,${}^{(5)}$
Michael C.~Ogilvie,${}^{(6)}$ R.~L.~Renken,${}^{(7)}$ Pietro~Rossi,${}^{(5)}$
D.~K.~Sinclair,${}^{(8)}$ R.~L.~Sugar,${}^{(9)}$
D.~Toussaint,${}^{(10)}$K.~C.~Wang${}^{(11)}$
\affil\vskip .10in
\centerline{${}^{(1)}$SCRI, Florida State University, Tallahassee, FL
32306-4052, USA}
\centerline{${}^{(2)}$University of Colorado, Boulder, CO 80309, USA}
\centerline{${}^{(3)}$Indiana University, Bloomington, IN 47405, USA}
\centerline{${}^{(4)}$University of Illinois, Urbana, IL 61801, USA}
\centerline{${}^{(5)}$Thinking Machines Corporation, Cambridge, MA 02139, USA}
\centerline{${}^{(6)}$Washington University, St.~Louis, MO 63130, USA}
\centerline{${}^{(7)}$University of Central Florida, Orlando, FL 32816, USA}
\centerline{${}^{(8)}$Argonne National Laboratory, Argonne, IL 60439, USA}
\centerline{${}^{(9)}$University of California, Santa Barbara, CA 93106, USA}
\centerline{${}^{(10)}$University of Arizona, Tucson, AZ 85721, USA}
\centerline{${}^{(11)}$University of New South Wales, Kensington, NSW 2203,
Australia}
\goodbreak
\if \preprint Y \preprintno{FSU-SCRI-92-08}\fi
\if \preprint Y \preprintno{COLO-HEP-304}\fi
\abstract
We have measured some simple matrix elements for pseudoscalar
and vector mesons made of Wilson valence quarks
and   staggered sea quarks at $\beta=5.6$
at sea quark masses $am_q=0.01$ and 0.025.
Our measurements include the decay constants of pseudoscalars
(including $f_D$), the
wave function at the origin  (or decay constant)
of vector mesons, and the calculation of
quark masses from current algebra.
The effects of sea quarks on the simulations are small.
We make comparisons to quenched simulations at similar values of the
lattice spacing ($1/a \simeq 2$ GeV.)
\endtitlepage
\head{I. Introduction}
We have been engaged in an extended program of calculation of the masses
and other parameters of the light hadrons in simulations which include
the effects of two flavors of light dynamical quarks\rlap.\refto{HEMCGC,LATEST}
  These quarks
are realized on the lattice as staggered fermions. We have carried
out simulations with lattice valence quarks in both the staggered
and Wilson formulations.  In this paper we compute simple matrix elements
 using valence Wilson quarks. Most previous work
with Wilson valence quarks has been
done in the quenched approximation. The quenched approximation is uncontrolled,
and one would like to know the magnitude of the effects of sea quarks on matrix
elements calculated in this approximation.
We can do this since we have performed simulations with two masses
of sea quarks. We can also compare our results to published results done in
the quenched approximation at equivalent values of the lattice spacing.
 Including the effects of staggered sea
quarks is computationally less intense than using Wilson sea quarks,
and we consider that mixing the two realizations is not inappropriate
 for a first round  of numerical simulations.

We have  measured matrix elements of the form $\langle h |A_\mu |0 \rangle$
and $\langle h |V_\mu |0 \rangle$ where $A_\mu$ is an axial current and $V_\mu$
is a vector current.
Physically, these quantities parameterize the
decay constants of pseudoscalar and vector mesons. Valence quark masses range
from very light through the charmed quark mass. In particular, we will present
a prediction for the decay constant of pseudoscalars containing a
charmed quark.
 As a byproduct of the axial current measurement, we
measure a quantity which is proportional to the valence quark mass $m_q$.
In addition, we determine the renormalization factors (or ratios of
renormalization factors) from various definitions of vector and
axial currents, which can then be compared to perturbation theory and to
other simulations.

The lattices we use are the ones we generated in our most recent round
of simulations\refto{LATEST}
on $16^3 \times 32$ lattices at a lattice coupling of $\beta=5.6$.
The simulations include two flavors of dynamical staggered fermions; the
dynamical fermion masses are $am_q=0.01$ and $am_q=0.025$.

We should warn the reader that these simulations are performed at
values of the lattice spacing which are quite a bit greater than
those used by present-day quenched simulations. Depending on the particle
whose mass is used to set the lattice spacing, our lattice
spacing $a$ lies in the range $1/a= 1700$ to 2100 MeV. Thus we expect
considerable contamination from lattice artifacts.  Probably the
most reasonable quenched approximation data sets with which to compare to are
ones taken at $\beta \leq 6.0$, since they are thought to have
a similar lattice spacing. (For example, the APE collaboration\refto{APE}
has determined a lattice spacing from quenched Wilson spectroscopy of
$1/a=2132$ MeV, by extrapolating the rho mass to zero valence quark mass,
or to the critical hopping parameter, $\kappa_c$.)
At the parameter values of these simulations spectroscopy is essentially
identical to that from quenched simulations.

The outline of the paper is as follows:
In Section II we describe the lattice simulations and the data set.
In Section III we describe the methodology for extracting matrix elements
from the data.
In Section IV we describe the program for converting lattice numbers
to continuum numbers and compare some results to perturbation theory.
Sections  V, VI and VII describe the vector decay constant, quark masses,
and the pseudoscalar decay constant. A few conclusions are in Section VIII.

\head{II. The simulations}
\subhead{A. Numerics}
Our simulations were performed on the Connection Machine CM-2 located
at the Supercomputer Computations Research Institute at Florida State
University.

We carried out simulations with two flavors of dynamical staggered
quarks using the Hybrid Molecular Dynamics algorithm\rlap.\refto{HMD}
The lattice size is $16^3 \times 32$ sites and the lattice coupling
$\beta=5.6$. The dynamical quark mass is $am_q=0.01$ and
0.025. The total simulation length was 2000 simulation time units
(with the normalization of Ref. \cite{HEMCGC}) at each quark mass value.
  We recorded lattices
for the reconstruction of spectroscopy every 20 HMD time units, for a total
of 100 lattices at each mass value.

We computed spectroscopy with staggered sea quarks at six values of the
Wilson quark hopping parameter: $\kappa=0.1600$, 0.1585, 0.1565, 0.1525,
0.1410, and 0.1320. The first three values are rather light quarks
(the pseudoscalar mass in lattice units
 ranges from about 0.25 to 0.45) and the other
three values correspond to heavy quarks (pseudoscalar masses of
 0.65 to 1.5). We computed properties
 of mesons with all possible combinations of
quark and antiquark mass; this will allow us to study matrix elements
of strange and charm
mesons.
 We used periodic
boundary conditions in all four directions of the lattice.
We fix gauge in each configuration in the data set to lattice Coulomb gauge
using an
overrelaxation algorithm\refto{OVERRELAX}.
Our inversion technique  is  conjugate gradient with preconditioning via
ILU decomposition by checkerboards\rlap.\refto{DEGRANDILU}
We used a fast matrix inverter written in CMIS
(Connection Machine Instruction Set)\refto{LIU}.

\subhead{B. Interpolating fields}
Matrix elements are determined from correlation functions such as
  $$C_{ij}(\vec k=0,t) = \sum_{\vec x}
\langle O_i(\vec x,t)O_j(\vec 0,t=0) \rangle . \eqno(2.1)$$
Just as in the case of spectroscopy, a good interpolating field is necessary
so that the correlator is dominated by the lightest state in its channel
at small times separation.  We have chosen to use an interpolating field
which is separable in the quark coordinates and extended in the coordinates of
either quark:
$$O_1(\vec x,t) = \sum_{y_1,y_2}\phi_1(\vec y_1 - \vec x)
 \phi_2(\vec y_2 - \vec x) c_q(\vec y_1,t)^\dagger \Gamma
 c_{\bar q}(\vec y_2,t)^\dagger . \eqno(2.2)$$
Here $c_i(\vec y,t)^\dagger$ are creation operators for quark and
antiquark, $\Gamma$ the appropriate Dirac matrix, and we have suppressed all
color and spin indices.
Since the operator is separable the individual $\phi$ terms are sources for
calculation of quark propagators.  We take $\phi(\vec x)$ to be
a Gaussian centered around the origin:
$$\phi(\vec x) = \exp(-(|\vec x|/r_0)^2) . \eqno(2.3)$$
The parameter $r_0$ can be chosen to give an optimal overlap  with the ground
state.

We  need zero spatial momentum correlation functions with $O_1$ also
as sink
$$\eqalign{
C_{11}(\vec k=0,t) = & ~ \sum_{\vec x}
\langle O_1(\vec x,t)O_1(\vec 0,0) \rangle \cr
 = & ~ \sum_{\vec x} \langle
 \sum_{y_1,y_2} \phi_1(\vec y_1 - \vec x) \phi_2(\vec y_2 - \vec x)
 \Gamma^\dagger G_q(\vec y_1,t; \phi_1, t=0)
\Gamma
 G_{\bar q}(\vec y_2,t; \phi_2, t=0) \rangle . \cr } \eqno(2.4) $$
Here $G_i(\vec y,t; \phi, t=0) = \sum_{\vec z} G_i(\vec y,t; \vec z, t=0)
\phi(\vec z)$ is obtained as the inverse of the fermion matrix on a source
$\phi(\vec z)$ on time slice $t=0$. Going to Fourier transforms we can
write this as
  $$C_{11}(\vec k=0,t) = \sum_{\vec p}
\langle \tilde \phi_1(\vec p) \tilde \phi_2(- \vec p)
 \Gamma^\dagger \tilde G_q(- \vec p,t; \phi_1, t=0) \Gamma
 \tilde G_{\bar q}(\vec p,t; \phi_2, t=0) \rangle, \eqno(2.5) $$
requiring, as for local sinks, only one sum over a time slice, plus the
cost of taking the Fourier transforms $\tilde G_i$ and $\tilde \phi_j$. The
latter can be precomputed once for the whole simulation. Using FFT's this
computation becomes practical. For our six $\kappa$ values combined, it took
only about $25\%$ longer than a computation using point or wall sources and
sinks.

\if\preprint Y \psoddfigure  486 466 -9 {Figure 1} {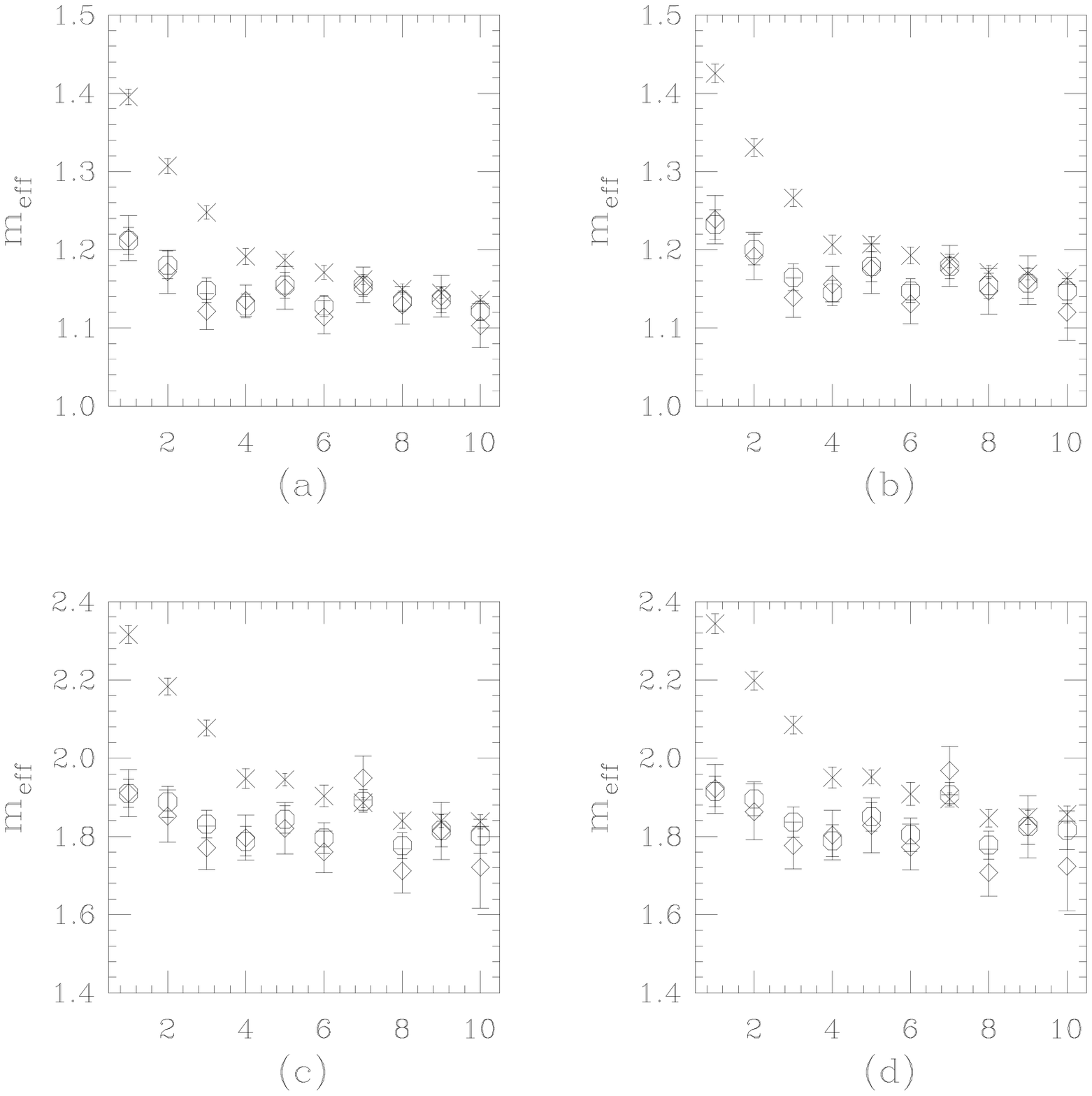} {
Effective masses for $\kappa=0.1410$ as a function of $r_0$: cross
$r_0=1$, octagon $r_0=2.5$, diamond $r_0=4.0$. Figures are (a) pion,
(b) rho, (c) nucleon, (d) delta.
}\fi
\if\preprint Y \psoddfigure  486 466 -9  {Figure 2} {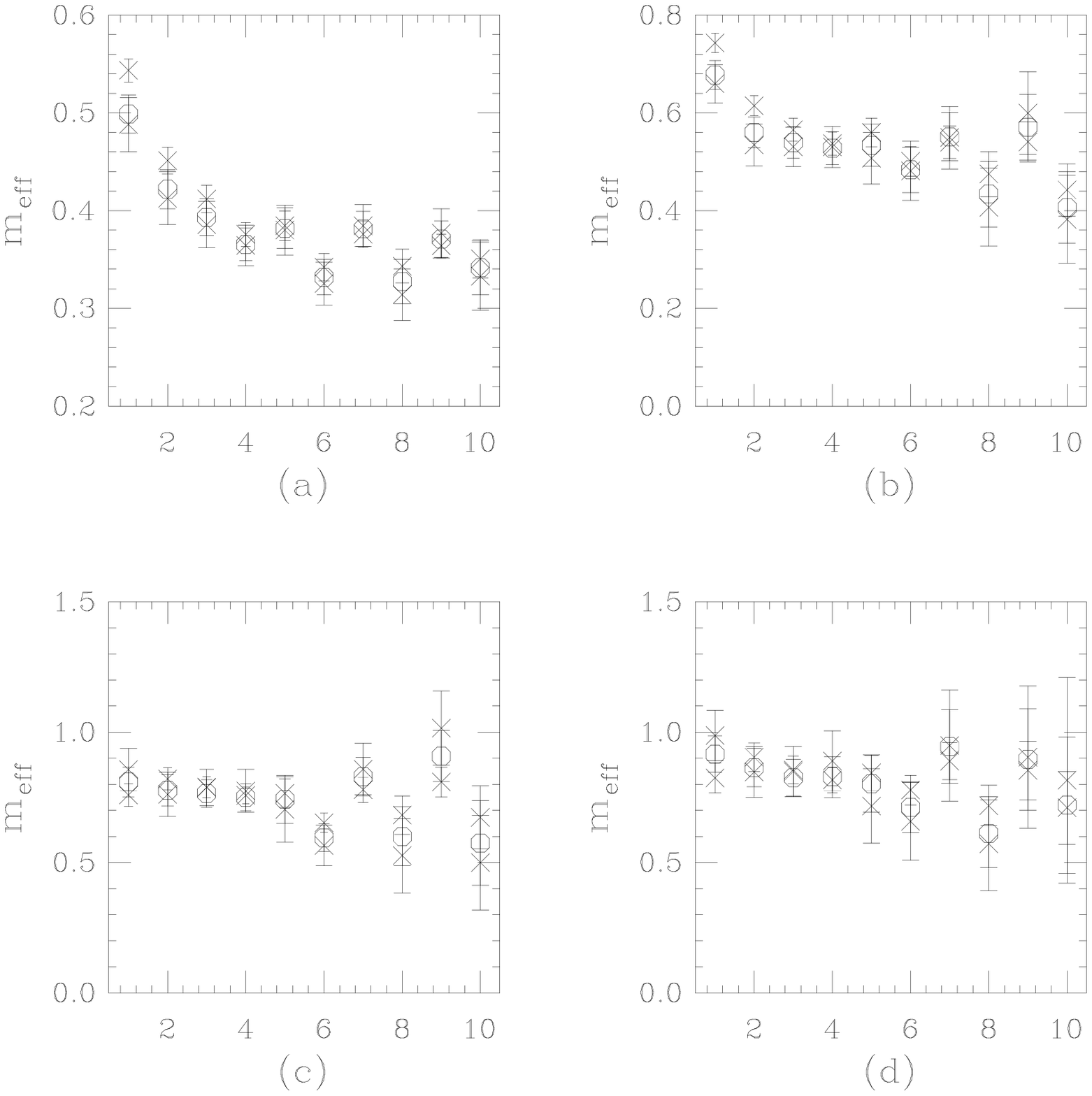} {
Effective masses for $\kappa=0.1585$  as a function of $r_0$: cross
$r_0=3.0$, octagon $r_0=4.0$, diamond $r_0=5.0$.
}\fi
We performed a small test to determine reasonable $r_0$'s:
we took 24 lattices and calculated hadronic spectroscopy at
two hopping parameter values  $\kappa=0.1410$ amd 0.1585,
at sea quark mass 0.025,
and three values of $r_0$ at each $\kappa$.
We show effective mass plots for hadrons made of degenerate heavy quarks and
degenerate light quarks in Figs. 1-2. While the light quark spectroscopy
was not very sensitive to $r_0$, the heavy quark spectroscopy was: a larger
$r_0$ was optimal. Following these tests, we chose the following
$r_0$ values for each $\kappa$ (shown as ($\kappa$, $r_0$):
(0.1320, 2.5),
(0.1410, 3.0),
(0.1525, 3.5),
(0.1565, 4.0),
(0.1585, 4.5), and
(0.1600, 5.0).
On the full data set we checked all the hadronic spectroscopy and
found agreement between results using these sources with our earlier
work using  a ``wall'' (uniform) source.

\head{III. General Formalism for Lattice Matrix Elements}

All  measurements  of matrix elements
involve determining ratios of correlation functions.
We have two generic classes of correlators: ones in which
the two operators are different (one of the operators
is to be determined and the other is not), $C_{12}$ of Eqn. (2.1),
and a class where both operators are identical to one of the operators in
the proceeding equation, $C_{11}(t)$.
For example, $O_1$ could be a Gaussian interpolating field and $O_2$ could
be a current.

Passing to momentum space and inserting
a complete set of relativistically normalized states,
 we have
   $$C_{ij}(t) =  \sum_n  {1 \over 2\mu_n}
 \langle 0 | O_i|n(k=0) \rangle \langle n(k=0)| O_j
 |0 \rangle  e^{-\mu_n t}. \eqno(3.1)$$
At large $t$ only one state (of mass $\mu$)
 should dominate the sum, and we expect to see
  $$C_{12}(t) =   {1 \over 2 \mu} e^{-\mu t} \langle 0
| O_1|h \rangle \langle h| O_2 |0 \rangle \eqno(3.2)$$
and
  $$C_{11}(t) =   {1 \over 2 \mu} e^{-\mu t} \langle 0
| O_1|h \rangle \langle h| O_1 |0 \rangle. \eqno(3.3)$$

Of course, it is incorrect to compute $ \langle h| O_2 |0 \rangle$
by directly taking ratios of $C_{12}$ to $C_{11}$.  The correct way to
analyze the data, which produces a meaningful $\chi^2$, is to do a  three
parameter
($\mu$,  $\langle 0| O_1|h \rangle $,$\langle h| O_2 |0 \rangle$)
simultaneous fit to both data sets which includes the full
 correlation matrix.
This is how we analyze all our data. This fitting method also allows us
to quote a meaningful confidence level for a fit.
 Reference \cite{DOUGFIT}
discusses this fitting procedure in detail.
A typical set of correlators and their fits are shown in Fig. 3.
\if\preprint Y \psfigure  252 108 {Figure 3} {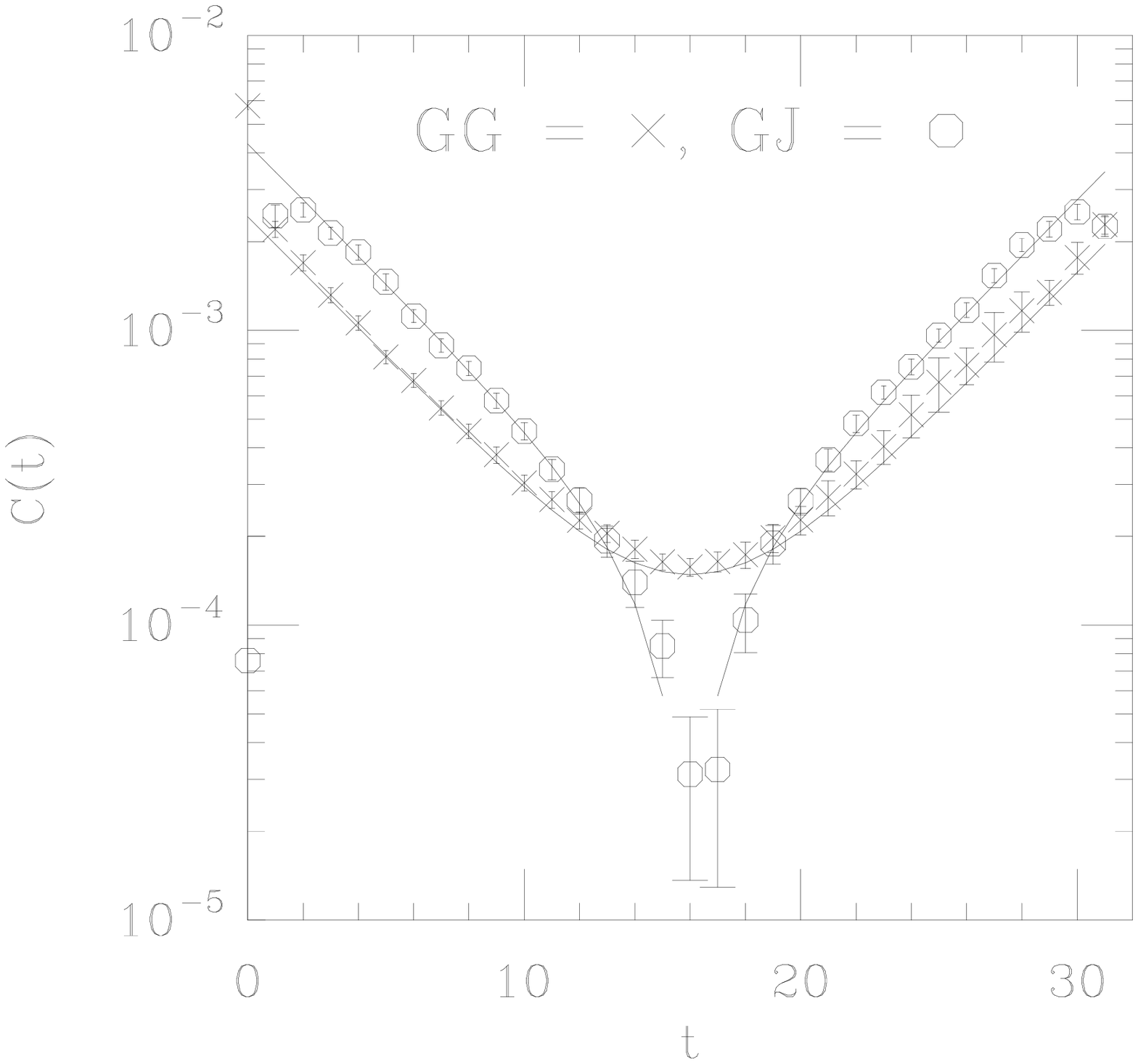} {
Data, and results of a three parameter correlated fit, to
 Gaussian source and sink (crosses), and to a Gaussian source
and  local axial current sink ($\bar \psi \gamma_0 \gamma_5 \psi$) (octagons),
for $am_q=0.01$ sea quark mass and $\kappa=0.1600$ valence quarks.
The second correlator is antiperiodic and its absolute value is shown.
}\fi

\if\preprint Y \psfigure  252 108 {Figure 4} {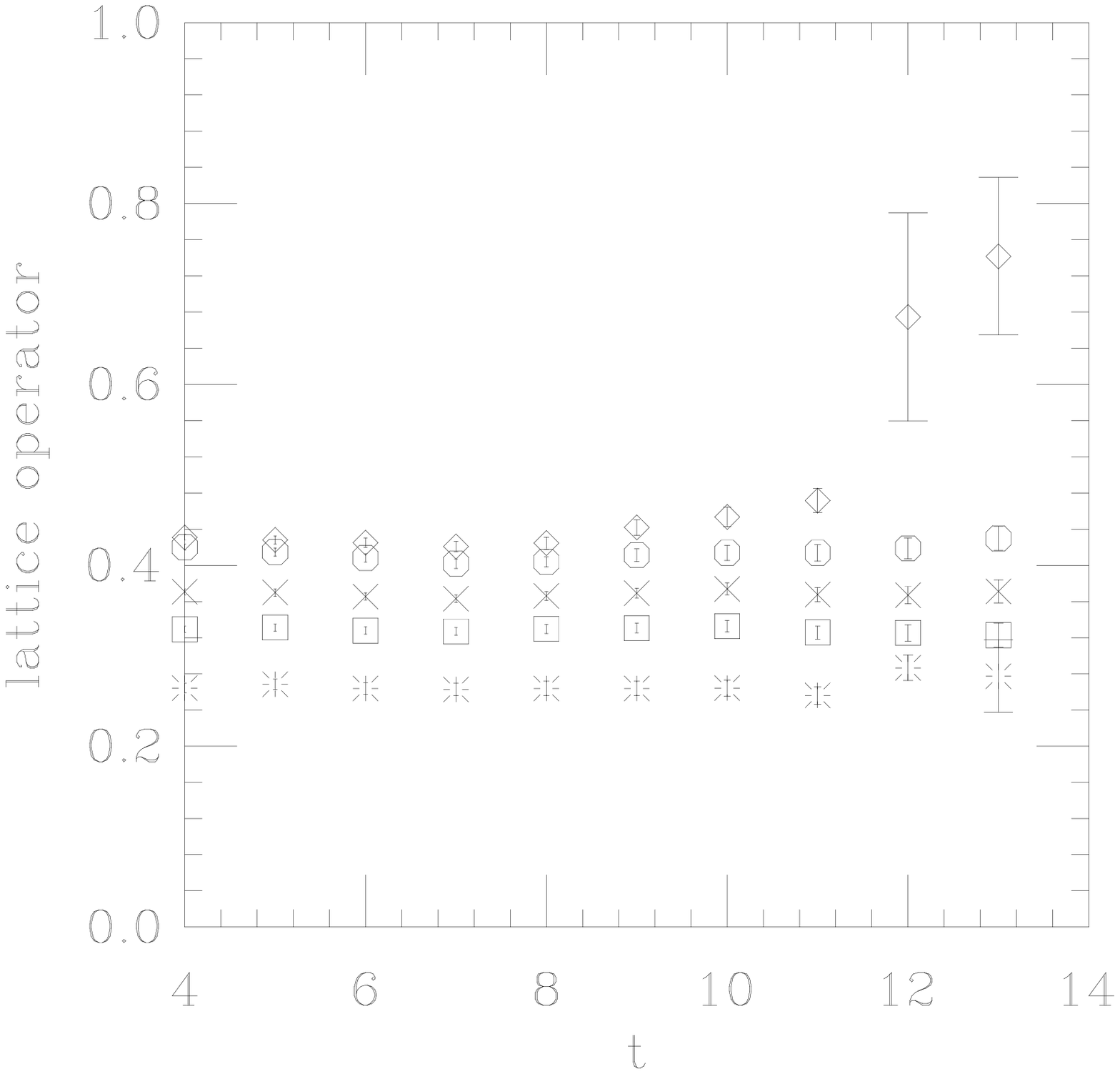} {
Results of fits to the lattice matrix element
$\langle 0 | \bar \psi \gamma_0 \gamma_5 \psi | P\rangle$
over the range $t$ to $t_{max}=15$,
from three parameter fits to two propagators.
The diamond, octagon, cross, and square are for heavy quark-light quark
pseudoscalars with heavy $\kappa=$ 0.1320, 0.1410, 0.1525 and 0.1565,
and the burst is for a meson made of two light quarks.
In all cases the light quark hopping parameter is $\kappa=0.1600$.
}\fi

One feature shown by all the operators we studied was the very poor
confidence level of fits to operators containing one (or two) heavy
quarks, especially the $\kappa=0.1320$ quark.
We saw this same behavior in our spectroscopy in Ref. \cite{LATEST}.
The culprit is the correlator with a local current at one end. While the
extended operator has reasonable overlap on the ground state, the local
current does not. In other words, $J(x,t)|0\rangle$ creates excited state
mesons much more readily than it creates ground states. Thus we do not see
a plateau in the effective mass of the state. The
operator itself drifts with fitting range, too.  The same problem has crippled
other calculations of heavy-light matrix elements\rlap.\refto{BADHEAVY}
An example of this behavior is shown for the local axial current matrix
element,
in Fig. 4.

Since we have so many  different operators and quark masses, it is
necessary to give some general rule for selecting the best fit
value to present in a figure or table.
In selecting the distance range to be used in the fitting, we have tried
to be systematic.  We somewhat arbitrarily choose the best fitting range
as the range which maximizes the
confidence level of the fit (to emphasize good fits) times the
number of degrees of freedom (to emphasize fits over big distance
ranges) divided by the statistical error on the mass (to emphasize fits with
small errors). We typically restrict this selection to fits beginning no
more than 11 or 12 timeslices from the origin.
This was the method we used to select the best mass in our earlier
work\rlap.\refto{HEMCGC}

When we have to extrapolate a quantity (for example, to zero quark mass),
we combine correlated fits with jackknife averages. We perform a series of
fits where ten successive lattices are removed from the data set, select the
best fit value of the matrix element using the procedure described in the
last paragraph, extrapolate the desired quantity for the subset of data,
and then perform a jackknife average of the extrapolated quantity.

\head{IV. From lattice to continuum numbers}
Next we must discuss how we convert lattice measurements into predictions
of continuum observables.  There are three complications: First, dimensionful
observables require a lattice spacing to set the scale. Since the simulations
do not show proper mass ratios for hadrons at better than the fifteen per
cent level, we cannot claim to know the lattice spacing better than this.
Second, lattice operators require perturbative
renormalizations to convert them to continuum.
Third, there are additional corrections ($O(a)$ corrections) which arise
because the lattice operators themselves differ from the continuum
operators by terms proportional to the lattice spacing $a$.
Thus, the continuum matrix element is related to the lattice one by
$$\langle f|O^{cont}(\mu)|i\rangle
 = a^D(Z(\mu a, g(a))\langle f|O^{latt}(a)\rangle + O(a) + O(g^2a) + \dots).
\eqno(4.1)$$
($D$ is the engineering dimension of the matrix element.)
A symptom of the presence of $O(a)$ contributions is an apparent dependence
of the $Z$ parameter on initial and final state.
These corrections can be ameliorated by using ``improved operators''
[\cite{IMPROVED,IMPROVEDB}]. This study used only unimproved operators,
however,
and the $O(a)$ corrections remain an unknown systematic.

The  $Z$-factors are a product of two terms.  The first is a
lattice to continuum regularization conversion factor
which in perturbation theory is a power series in $\alpha_s$.
The second term is an overall multiplication by a function of the hopping
parameter $\kappa$ which relates the field normalization for Wilson
fermions to the continuum fermion normalization.

There are two ways
to assign this normalization. The first is to relate the actions of
Wilson fermions to the continuum action. This gives a multiplication
factor of $\sqrt{2\kappa}$ as the conversion between Wilson and
continuum fermions.

The second way follows a recent paper by Lepage and
Mackenzie\refto{PETERPAUL}.  They suggest a program for calculation
of renormalization terms which has two parts: ``tadpole improved''
perturbation theory and a choice of subtraction scheme for perturbative
calculations which attempts to minimize higher order corrections.

Tadpole improved perturbation theory for Wilson fermions is based on the
observation that ordinary perturbation theory, based on the naive
expansion
$$U_\mu(x) = 1 + igaA_\mu(x) + \dots \eqno(4.2)$$
is a misleading  expansion in $g$. Higher order terms in the expansion of
the exponential contain higher powers of $agA_\mu$ and the ultraviolet
divergences from contractions of $A$'s cancel the extra $a$'s, leaving
these terms suppressed not by powers of $ga$, but only by powers of $g$.
These contributions are the QCD tadpole corrections.
Tadpoles renormalize the link operator (at least in smooth gauges), and so
Lepage and Mackenzie suggest writing
$$U_\mu(x) = u_0(1 + igaA_\mu(x) + \dots) \eqno(4.3)$$
where $u_0$ is chosen nonperturbatively. This suggests replacing
$U_\mu$ in the action by the combination $U_\mu/u_0$, so that the
tadpole-improved action for Wilson fermions is
$$S = \sum_n \bar \psi_n \psi_n -  \tilde \kappa \sum_{n \mu}(\bar \psi_n
(1 - \gamma_\mu) {{U_\mu(n)}\over u_0} \psi(n+ \mu) + \dots \eqno(4.4) $$
Here $\tilde \kappa = \kappa u_0$. At tree level $\tilde \kappa_c = 1/8$ and
one
might expect that $\kappa_c \simeq 1/(8u_0)$.

Now compute the field renormalization in the following way:
Compute the quark two-point functions $S_F(ip_0, \vec p =0)$. Near its
pole
$$S_f(ip_0) \simeq {{1+\gamma_0}\over 2}{1 \over {(p_0 - \mu)}}
{1\over{(1 - 6 \tilde \kappa)}} \eqno(4.5)$$
where the quark mass  $\mu$
solves $1 - 6 \tilde \kappa = 2 \tilde\kappa \exp(-\mu)$.
(This is the ``$\exp(ma)$ factor'' recently advocated by
Bernard, Labrenz, and Soni\refto{BLS} in their analysis of pseudoscalar
decay constants.)
This implies that
$\psi_{contin} = \sqrt{1-6 \tilde\kappa} \psi_{lat}$ or that the
overall field renormalization factor should be
$\sqrt{1 - 3\kappa/4\kappa_c}$.

This is different from the conventional $\sqrt{2 \kappa}$ field
renormalization. For our data, where at sea quark mass $am=0.01$,
$\kappa_c = 0.1610$, the square of the field renormalization  in this
prescription varies
from 0.385 to 0.255 as $\kappa$ varies from 0.1320 to 0.1600, while
the variation in the naive prescription is only 21 per cent, and the
 factors vary from 0.262 to 0.32.

To calculate tadpole-improved renormalizations of matrix elements
 for massless quarks,
Lepage and Mackenzie give the following prescription:
For local operators, replace the $\sqrt{2\kappa_c}$ field renormalization
by the perturbative expansion for $\kappa_c$ [\cite{GROOTHS}]
$$2\kappa_c = {1\over 4}(1 + 1.364 \alpha_s) \eqno(4.6)$$
so that a bilinear $\bar \psi \Gamma \psi$, with a perturbative
$Z$-factor of $(1 + A\alpha_s)2\kappa_c$
becomes
$(1 + (A+1.364)\alpha_s)1/4$
or at $\kappa \neq \kappa_c$
$(1 + (A+1.364)\alpha_s)(1 - 3\kappa/4\kappa_c)$.
This lowers the coefficient of $\alpha_s$ if $A$ is negative
(which it is for all bilinears we have seen).
Operators already containing a link ($\bar \psi \Gamma U \psi$)
automatically include the $u_0$ factor and the only change they require is
to convert $2\kappa_c$ into $1/4$.
Table I shows the renormalization factors for all the operators used in this
study. They are given in Refs. [\cite{PETERPAUL,GROOTHS,MANDZ}].

Lepage and Mackenzie suggest picking
 $u_0 = \langle {1\over 3}\Tr U_P\rangle^{1/4}$
as a nonperturbative definition of $u_0$ in terms of the measured
plaquette. In our simulations we found
$ \langle {1\over 3}\Tr U_P\rangle =0.56500(2)$ and  $0.56444(2)$
at $am_q=0.01$ and $0.025$ (note the tiny dependence on the quark mass).
This would give
$\kappa_c = 1/(8u_0) = 0.1442$ for both $am_q=0.01$ and $0.025$,
to be compared with $0.1610$ and $0.1613$, respectively.
The tadpole gives about half the observed renormalization in $\kappa_c$.

We will present results from both normalization schemes since most published
calculations of matrix elements use the conventional normalization.
This will also give the reader a feel for the magnitude of systematics
in lattice calculations which are not due to the simulations themselves.

Next one must define the coupling $\alpha_s$. Lepage and Mackenzie argue
that  the lattice coupling $\alpha_{latt} = 6/(4 \pi \beta)$
is a poor expansion parameter since with it coefficients of second order
corrections are large. They suggest using an alternative coupling defined
through the plaquette\refto{URS}
$$-\ln \langle {1 \over 3} \Tr U_P \rangle = 4.18879 \alpha_V(3.41/a)
\left\{1-(1.185 + 0.070 n_f) \alpha_V + O(\alpha_V^2) \right\} . \eqno(4.7)$$

{}From the measured
$\langle {1\over 3}\Tr U_P\rangle$ ($0.56500(2)$ and  $0.56444(2)$)
we obtain
$\alpha_V(3.41/a)=0.1785$ and $0.1790$ for $am_q=0.01$ and $0.025$.
These
numbers are probably given with too much accuracy. With the coefficient of
the $O(\alpha_V^2)$ correction in Eqn.~(4.7) assumed to be of order one, this
correction could easily change $\alpha_V(3.41/a)$ by $0.005$ or so.

Lepage and Mackenzie give an $O(\alpha_s)$ tadpole-improved perturbative
prediction
for $\kappa_c$ which works quite well for quenched simulations
for $\beta > 5.7$:
$${1 \over {2 \kappa_c}}= 4\langle {1\over 3}\Tr U_P\rangle^{1/4}
-1.268\alpha_V(1.03/a) \eqno(4.8)$$
The coupling is measured at lower $q$ since the UV-sensitive tadpoles
have been removed.
Using
$$ \alpha_V(q)^{-1} = \beta_0 \ln({q\over {\Lambda_V}})^2)
+ {\beta_1 \over \beta_0} \ln  \ln({q\over {\Lambda_V}})^2)) \eqno(4.9)$$
or
$$\Lambda_V/q = ({1\over {\beta_0 \alpha_V}})^{{\beta_1}/{2 \beta_0^2}}
\exp(-{1\over{2\beta_0 \alpha_V}}), \eqno(4.10)$$
where
$$\beta_0 = {{11 - {2\over 3}n_f}\over {4 \pi}} \eqno(4.11)$$
and
$$ \beta_1 ={{102 - {38 \over 3} n_f}\over {16 \pi^2}} \eqno(4.12)$$
are the first two coefficients of the beta function,
we run the $\alpha_V$'s down to the lower scale. We then find the perturbative
predictions of $\kappa_c=0.1624$ and $0.1626$ for $am_q=0.01$ and $0.025$
with $n_f=2$, to be contrasted with
the observed $0.1610$ or $0.1613$. Taking into account our estimated
uncertainty of $\alpha_V(3.41/a)$ of about $0.005$ which translates into an
uncertainty in $\kappa_c$ of about $0.001$ this is quite good agreement,
and is somewhat
better than the prediction of $0.1636$ which would obtain if $n_f=0$.

The perturbative formulae are actually valid for massless quarks.
Extrapolating $\langle {1\over 3}\Tr U_P\rangle$ linearly in $am_q$ to zero
quark mass, we then obtain $\alpha_V(3.41/a)=0.1782$ and $\kappa_c =
0.1623$. Given the uncertainty of about $0.001$ in this quantity, it is
again in quite good agreement with our extrapolated (again linearly in
$am_q$) $\kappa_c(am_q=0) = 0.1608(1)$. Along the way we find that
$a \Lambda_V = 0.201$ or with a nominal
lattice spacing $1/a \simeq 2.0(2)$ GeV
from our spectroscopy and the conversion of\refto{SCALES}
$$
{\Lambda_{\overline{MS}} \over \Lambda_V} = \exp \left\{ - {{31 N - 10 n_f}
\over {66 N - 12 n_f}} \right\} \eqno(4.13)
$$
for an $SU(N)$ gauge group with $n_f$ flavors of fermions that
$\Lambda_{\overline{MS}}=264(26)$ MeV. This is a lattice prediction which
includes the effects of two light fermions. The error here comes only from
the uncertainty in $a^{-1}$. The uncertainty in $\alpha_V(3.41/a)$
translates into an uncertainty of about $20$ MeV in
$\Lambda_{\overline{MS}}$.

We now begin a series of specific calculations of matrix elements.
When we speak of ``conventional'' calculations we mean that we
use $\sqrt{2\kappa}$ field normalizations and do not remove tadpole terms
from the perturbative $Z$ factors; however, we use $\alpha_V$ for all
coupling constants, but evaluated at a UV dominated scale,
$\alpha_V(3.41/a)=0.18$. For the tadpole improved calculations, the
coupling should be taken at a lower scale. Lepage and Mackenzie have not
analyzed the optimal scale for each operator. However it seems reasonable to
take a scale similar to the one used for the tadpole improved perturbative
estimation of $\kappa_c$, $\alpha_V(1.03/a)=0.31$. Since the perturbative
prediction for $\kappa_c$ are somewhat off, one might alternatively
estimate this coupling from Eq.~(4.7) using the measured values for both
plaquette and $\kappa_c$. Using the values extrapolated to zero mass for
these quantities we obtain $\alpha_V(1.03/a)=0.28$. We see that there is
about a 10\% uncertainty in the coupling constant needed for the
perturbative $Z$ factors, which implies an uncertainty of up to about 5\%
in the $Z$ factors. In the following we will use for the tadpole
improved analysis a coupling $\alpha_V=0.3$.

\head{V. Vector matrix elements}
We have measured matrix elements of three vector current operators,
the ``local'' vector current
$$V_\mu^l = \bar \psi \gamma_\mu \psi \eqno(5.1)$$
the ``nonlocal'' current
$$V_\mu^{nl} = {1\over 2}(\bar \psi \gamma_\mu U_\mu \psi + h.c.) \eqno(5.2)$$
and the conserved Wilson current
$$V_\mu^W = {1 \over 2}(\bar \psi ( U_\mu (\gamma_\mu-1) +
 U_\mu ^\dagger(\gamma_\mu+1) ) \psi). \eqno(5.3)$$
We extract the current matrix element from correlated fits to three
parameters of two propagators with the appropriate
operator as an interpolating field.

We choose to quote our vector current matrix elements through the
dimensionless parameter $f_V$
$$ Z_V  \langle V | V_\mu | 0 \rangle = {1 \over f_V} m_V^2 \epsilon_\mu.
\eqno(5.4)$$
In terms of this definition the width of a vector state (whose quarks have
a charge $e_Q$ in units of the electron's charge) to decay into
an $e^+ e^-$ pair is
$${1\over f_V^2} = {{3\Gamma(V \rightarrow e^+e^-)}\over
{4 \pi \alpha^2 e_Q^2m_V}} . \eqno(5.5)$$
This is the optimal  definition for a lattice calculation
 since there is no dependence on the lattice spacing.
The Wilson current is conserved but the other currents are multiplicatively
renormalized. The renormalization factors for all three currents are
shown in Table I.

How do the ratios of the $Z$-factors measured experimentally
compare with the perturbative predictions?  We measure these factors
by doing a correlated fit to the Wilson current and to one of the
nonconserved currents. This will extract the
$(1 + A \alpha_V)$ part of $Z$.  Our results are shown in Fig. 5 and
Tables II-V.
We find first that the Z-factors show only a few per cent variation
with $\kappa$, justifying the contention that the $\kappa$ dependence
 factorizes.  However, the perturbative formula predicts a ratio of
$$R^{l,w} = {{\langle 0|J^w|V\rangle}\over{\langle 0|J^l|V\rangle}}=
 Z^l/Z^W=0.86$$
and
$$R^{nl,w} = {{\langle 0|J^w|V\rangle}\over{\langle 0|J^{nl}|V\rangle}}=
 Z^{nl}/Z^W=0.83,$$
 quite a bit larger  than our
data. Our data resembles old results from quenched simulations. For
example, Martinelli and Maiani\refto{MANDM} measured
 $R^{l,w}=0.57(2)$ and $R^{nl,w}= 0.69(2)$
for these ratios in quenched $\beta=6.0$ simulations.

Results for the conserved current without any $Z$ factors are
shown in Tables VI-VII.
Generically all fits to mesons containing at least one of the heaviest quark
have low confidence levels.

Next, we wish to compare the measured $f_V$ to experiment.
 We first choose to
use the conserved Wilson current in this comparison. Our results
are shown as a function of the
   dimensionless mass ratio
$m_P^2/m_V^2$ in Fig. 6.  The data undershoot the lightest vector mesons
by about twenty five per cent but appear to reproduce experiment
for the $\psi(3100)$.
 Some of these
shifts are known to be due to the use of ``unimproved''
 operators.
A calculation\refto{IMPROVEDB}
 at $\beta=6.0$ shows that a conserved and improved
($CI$) vector current would have a matrix element ratio
$ = {{\langle 0|J^{CI}|V\rangle}/{\langle 0|J^{w}|V\rangle}}=1.20.$
It also finds $R^{l,w}=0.62$.

To compare with other simulations, we use a conventional normalization
for our data. We show in Fig. 7
 the conserved vector current and the local vector current.
We rescale the latter by the Martinelli-Maiani factor of 0.57.
Quenched $\beta=6.0$ results from Daniels, et. al., Ref. \cite{GUPTA},
and from APE, Ref.   \cite{APE},
 also use local currents and the same Z-factor
and are also shown.
(If we assumed $Z_V$ could be written as
$1 + A \alpha_s$ and just rescaled by the naive ratio of
couplings 5.6/6.0, it would change to 0.54, an invisible
variation on the plot.)
Our data is quite similar to the quenched results.
Again, dependence of the matrix element
on the sea quark mass is small.

Figs. 6 and 7 show that the same lattice data can give quite different
results depending on how it is converted to continuum numbers, even though
the calculated quantity is dimensionless.
 For the vector operators,
the discrepancy is largest for small quark mass.

\if\preprint Y \psoddfigure  486 466 -9 {Figure 5} {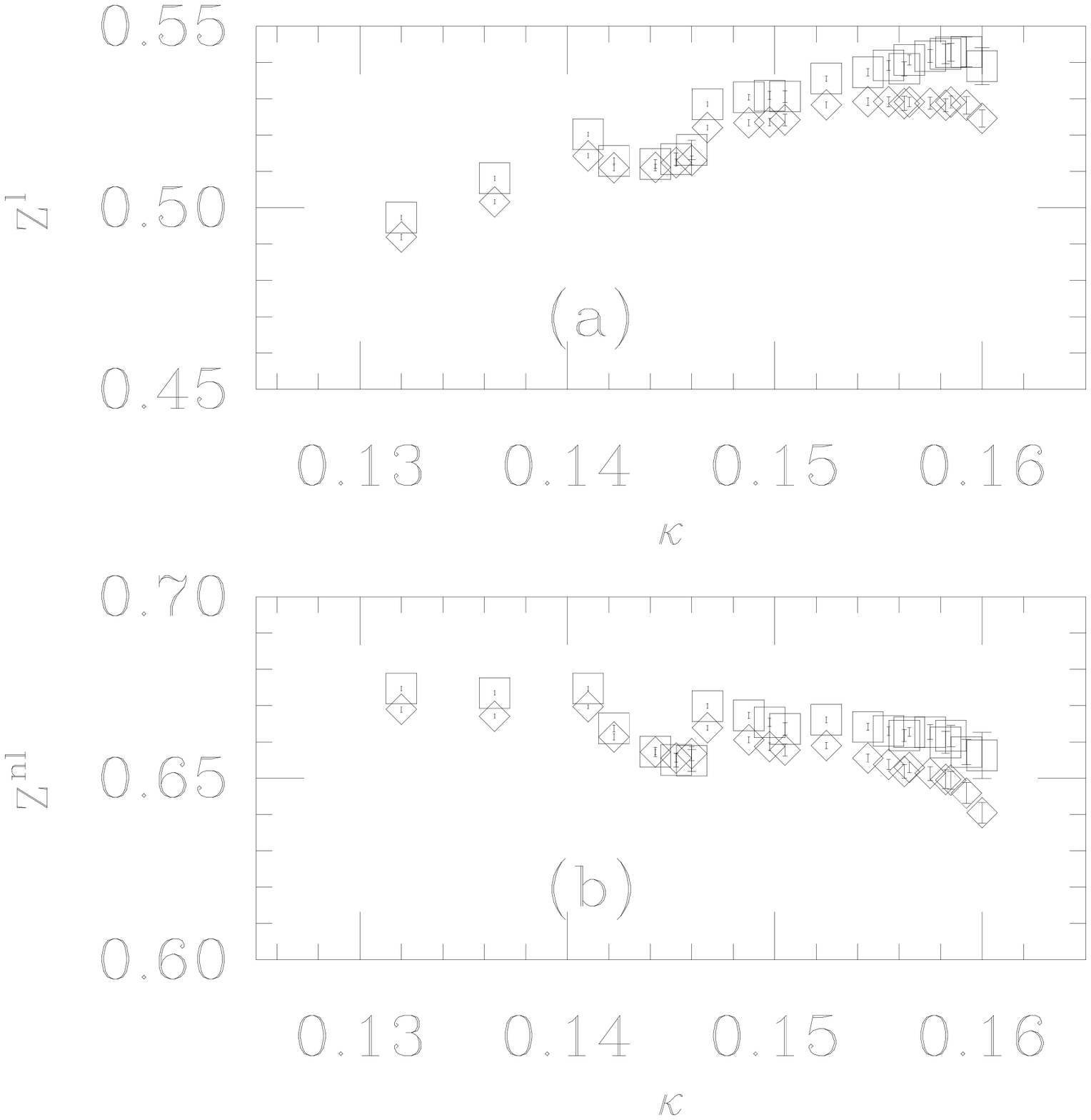} {
Ratios of renormalization factors for (a) local and (b) nonlocal vector
currents to the conserved current.
Results from simulations with sea quark mass $am_q=0.01$ are shown in
squares, and for sea quark mass $am_q=0.025$  in diamonds.
}\fi

\if\preprint Y \psfigure  252 108 {Figure 6} {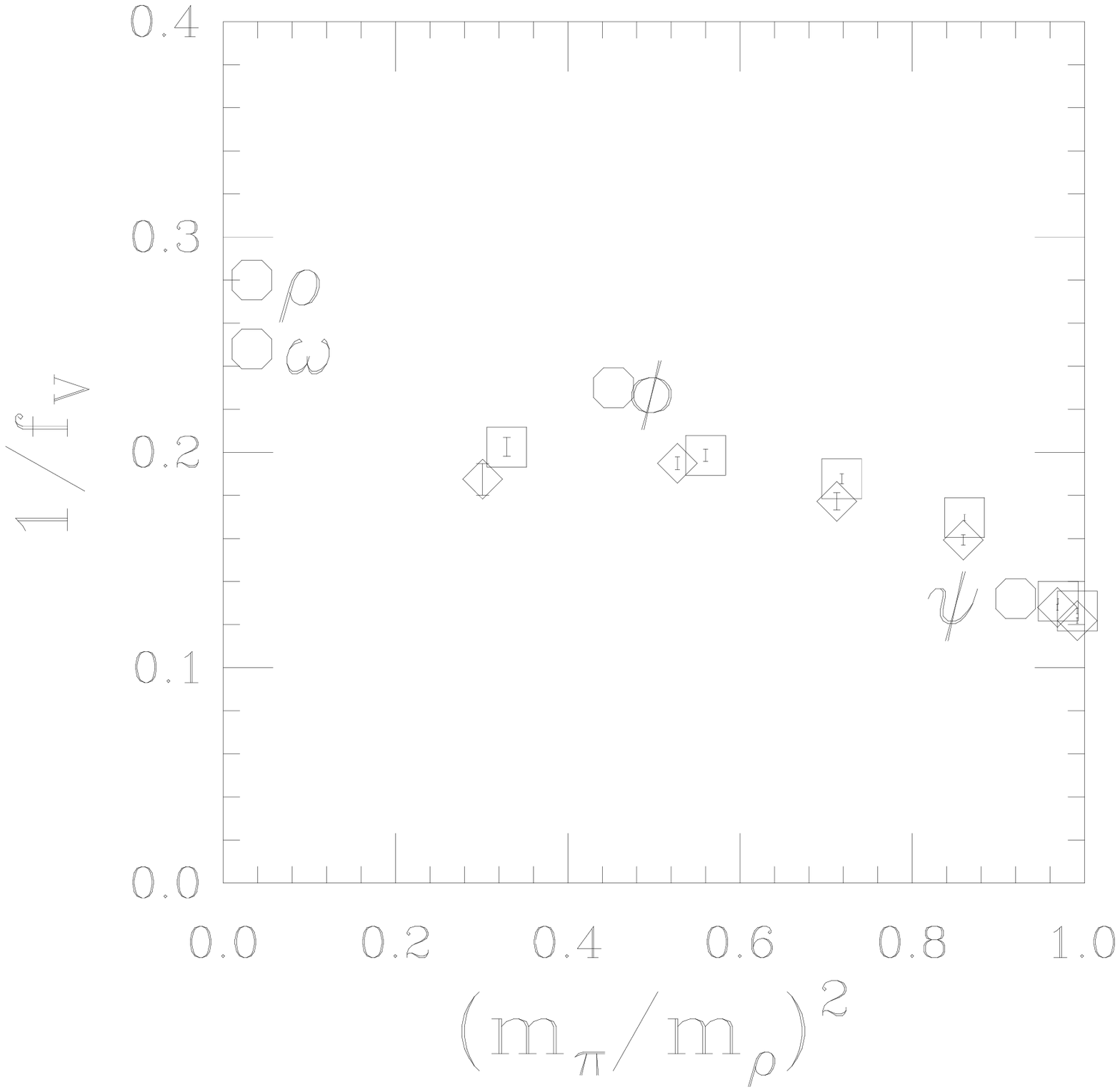} {
Lattice $1/f_V$ from the conserved  (Wilson)
vector current, as a function of  $(m_\pi/m_\rho)^2$,
using tadpole
improved perturbation
theory. The labeled points are physical particles
Results from simulations with sea quark mass $am_q=0.01$ are shown in
squares, and for sea quark mass $am_q=0.025$ in diamonds.
}\fi

\if\preprint Y \psfigure  252 108 {Figure 7} {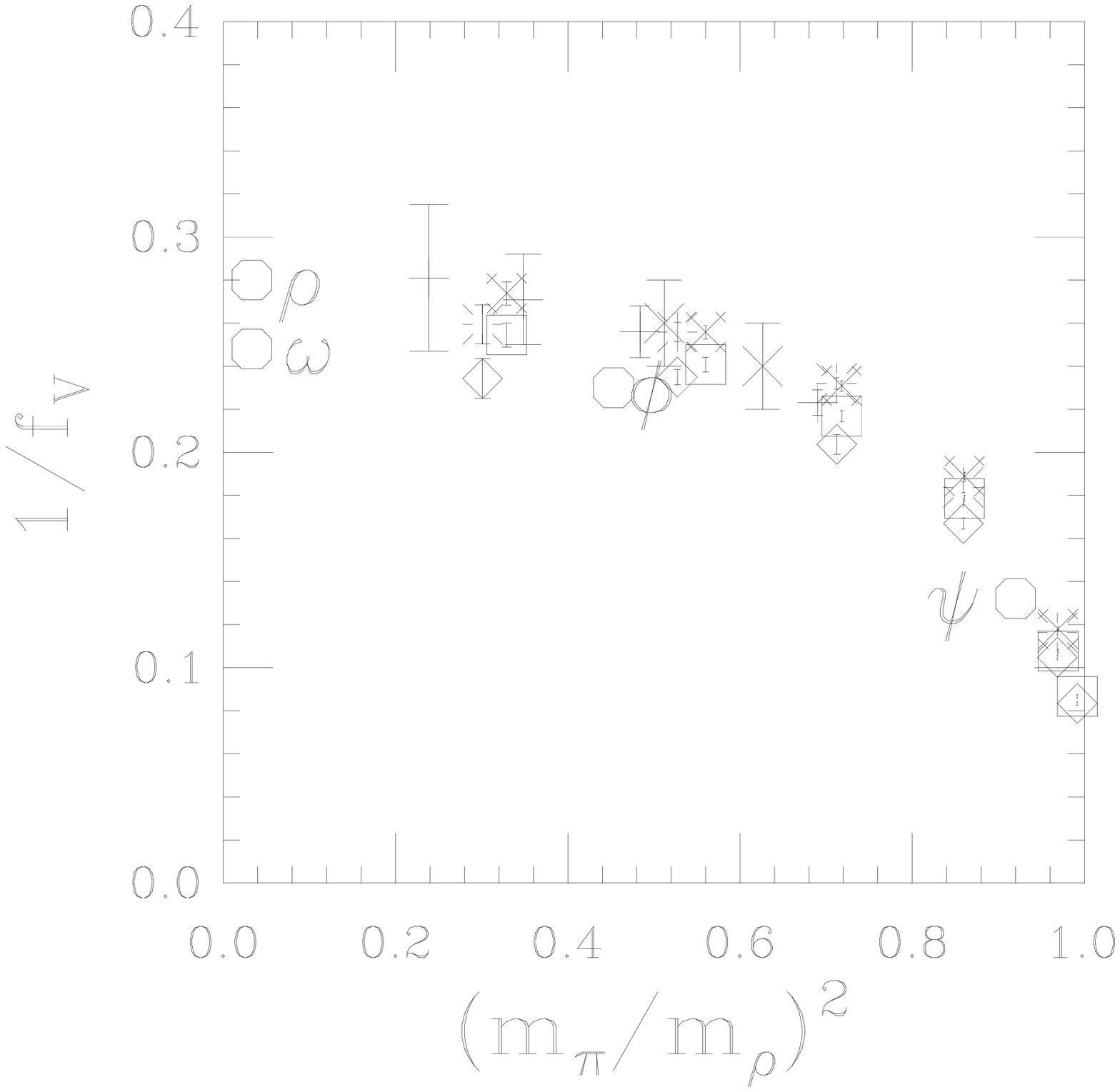} {
Lattice $1/f_V$ as a function of  $(m_\pi/m_\rho)^2$, using conventional
field normalization and perturbative corrections. The labeled points are
physical particles. Results for the conserved (Wilson) vector current
from simulations with sea quark mass $am_q=0.01$ are shown in
squares, and for sea quark mass $am_q=0.025$ in diamonds.
Results  for the local vector current, scaled by a phenomenological
$Z_V=0.57$
from simulations with sea quark mass $am_q=0.01$ are shown in
fancy crosses, and for sea quark mass $am_q=0.025$ in bursts.
Quenched $\beta=6.0$ from Daniels, et. al., Ref. \cite{GUPTA}, are crosses,
and APE results  Ref. \cite{APE} are pluses.
}\fi

\head{VI. Quark masses}
The basic relation  which gives us a quark mass is a
continuum current algebra relation
$$
\nabla_\mu \cdot \langle \bar\psi \gamma_5 \psi(0) \bar\psi\gamma_5 \gamma_\mu
\psi(x) \rangle =
2m_q \langle\bar\psi \gamma_5 \psi(0) \bar\psi\gamma_5 \psi(x) \rangle
\eqno(6.1)$$
If we  convert to lattice operators,
sum over spatial slices, and measure distance in the $t$ direction,
this becomes:
$$
Z_A{\partial\over{\partial t}} \sum_{x,y,z} \langle \bar\psi \gamma_5 \psi(0)
\bar\psi\gamma_5 \gamma_0 \psi(x) \rangle =
2a m_q Z_P\sum_{x,y,z} \langle\bar\psi \gamma_5 \psi(0) \bar\psi\gamma_5
\psi(x) \rangle. \eqno(6.2)$$
There are many possibilities for defining an axial current and for
defining the derivative operator\rlap.\refto{MANDM,IWASAKI}
  The most stable one we have
found\refto{DOUGFPI} is to say that since $A(z) \approx \sinh(m_\pi(t-N_t/2))$,
$$
{{\partial A(t)}\over{\partial t}} = m_\pi\cosh(m_\pi(t-N_t/2))
. \eqno(6.3)$$
Then we extract the quark mass by fitting
$$P(t) = Z(\exp(-m_\pi t) + \exp(-m_\pi (N_t -t) ) \eqno(6.4)$$
and
$$A(t) = {Z_P \over Z_A}
{{2m_q}\over m_\pi} Z(\exp(-(m_\pi t) - \exp(-m_\pi (N_t -t) ). \eqno(6.5)$$

We determine quark masses from both local and nonlocal axial
 currents. Our results are shown in Tables VIII-XI.
The reader should note that the best fits for mesons containing one or two of
the  heaviest quarks $\kappa=0.1320$ ) are unacceptably poor.
{}From the table the reader can convince him/herself that the average
quark mass interpolates quite well among the different $\kappa$ values--
that is, $m_q(\kappa_1+ \kappa_2) \simeq 1/2(m_q(\kappa_1) + m_q(\kappa_2))$.
As an alternate display, we show in Fig.~8
quark masses for all combinations of
quarks as a function of $1/\kappa_{ave} - 1/\kappa_c$
where $1/\kappa_{ave} = 0.5( 1/\kappa_1 + 1/\kappa_2)$.
 The data are remarkably linear.

We can compare our results to very simple theory.  In  the free field limit
the relation between the quark mass and the hopping parameter
for small quark mass is
$$ am_q =
{1 \over 2} ( {1\over \kappa } - {1\over \kappa_c })\eqno(6.6)$$
(with $\kappa_c=1/8$). From the discussion of Sec. III we have
$$ am_q = \log ( {{1 - 6 \tilde \kappa}\over{2 \tilde \kappa}} )  \eqno(6.7)$$
with $\tilde \kappa = \kappa/(8\kappa_c)$.
We can take $\kappa_c$ from our extrapolation of the pion mass
 and plot
Eq. (6.6) and (6.7) on Fig. 8.  Both curves provide a good representation
of the data. Note that this implies that the renormalization
of the prefactor (1/2) is much smaller than the renormalization of $\kappa_c$,
especially for the  quark mass extracted from the local axial current.

\if\preprint Y \psoddfigure  486 466 -9 {Figure 8} {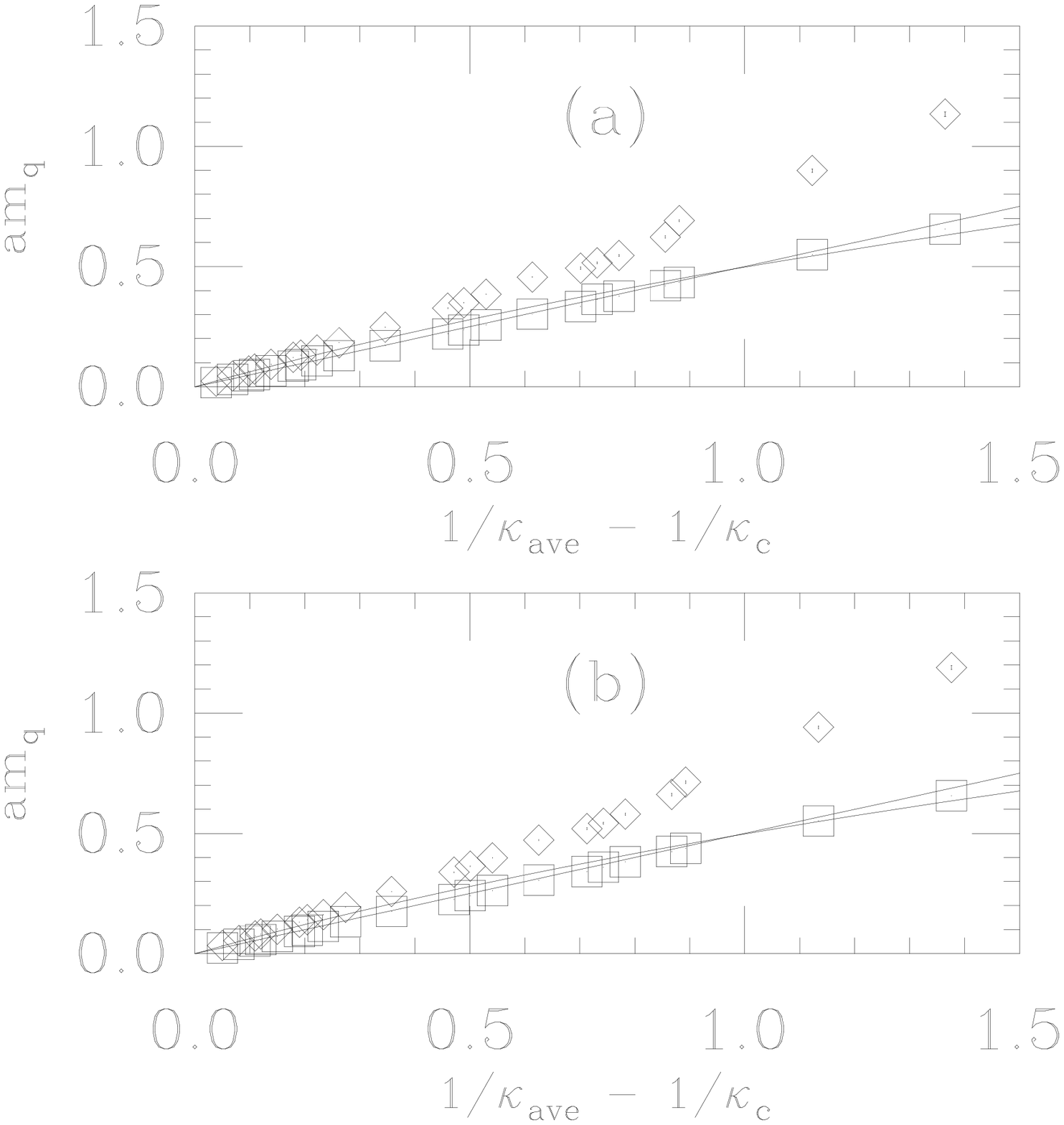} {
Quark mass in lattice units as a function of   $1/\kappa_{ave} - 1/\kappa_c$.
Results for the local axial current are shown in
squares, and for the nonlocal axial current in diamonds.
The lines show predictions of Eqns. (6.6) and (6.7).
(a) Sea quark mass $am_q=0.01$, (b) sea quark mass $am_q=0.025$.
}\fi

One should be able to measure $\kappa_c$ from the point where the
quark mass vanishes.  This should serve as a reasonably independent check
of the calculation of $\kappa_c$ from the vanishing of the pion mass.
(It is not completely independent, since the same lattices and some of the same
operators are used in both extrapolations).
We previously reported $\kappa_c=0.1610(1)$ for $am_q=0.01$
and $am_q=0.1613(1)$ for $am_q=0.025$.  The latter value was different
from that found from simulations on $12^4$ lattices, and it is important
that it be re-checked.
We do this as follows: We take only the six lightest $(\kappa_1,\kappa_2)$
combinations (all combinations of the three lightest $\kappa$ values).
We take
$$am_q = A(1/\kappa - 1/\kappa_c)  \eqno(6.8)$$
as  a model for the dependence of the quark mass on $\kappa$.
All the data comes from the same set of lattices, and so is strongly
correlated. In order to assign an uncertainty to the fit parameters we perform
a jackknife fit, dropping ten successive lattices from each of ten
subsets (containing ninety lattices) from the data, fitting each subset,
extracting a $\kappa_c$, and averaging.
We set all $Z$ factors to unity; we are extrapolating the quark
 mass in lattice
units with a lattice cutoff to zero.
We report the results of these fits in Table XII.
The critical hopping parameter determined from the quark mass appears to
be consistent with the value determined from the vanishing of the pion mass.

Finally, we can compare these Wilson quark masses to the quark masses
of the staggered sea quarks. One way to do this is to match pion masses.
The $am_q=0.025$ staggered pion has a mass in lattice units of about 0.42,
and the $am_q=0.01$ staggered pion has a mass of about 0.26. The former
is about halfway between the masses of pseudoscalars with valence
Wilson quarks of $\kappa=0.1565$ and 0.1585, 0.47 and 0.36 (at sea quark
mass $am_q=0.025$. The quark mass at those hopping parameters is 0.08
and 0.05. The $am_q=0.01$ pion lies between the $\kappa=0.1600$ and 0.1585
pions (0.21 and 0.33). However, the quark masses at those hopping
parameters are 0.04 to 0.02. Thus the quark mass determined with the method
of this section is much greater than the staggered quark mass at similar
values of the pion mass.

\head{VII. Axial matrix elements}
We  measured matrix elements of two axial current operators,
the local current
$$A_0^{loc} = \bar \psi \gamma_0 \gamma_5 \psi \eqno(7.1)$$
and the nonlocal operator
$$A_0^{nl} = {1\over 2}(\bar \psi  U_0 \gamma_0 \gamma_5 \psi + h.c.).
\eqno(7.2)$$
We  used two source spinor combinations for the local operator:
an interpolating field $\bar \psi \gamma_5 \psi$ and
$\bar \psi \gamma_0 \gamma_5 \psi$. Only the first source was used
with the nonlocal operator. This gives us a check of possible
sensitivity to the source of our measurements.
We quote answers in terms of the lattice pseudoscalar decay constant
    $$ Z_A  \langle 0| A_0|P  \rangle =f_P^L  m_P. \eqno(7.3)$$

Our values for $f_P^L$ are given in Tables XIII-XIV,
and shown in Fig. 9, for local and nonlocal operators.
The figure includes all $Z$ factors
from a tadpole improved calculation and lacks only a lattice spacing to
show continuum numbers. The tables include no Z-factors, to
facilitate comparisons with other simulations (they are $f_P^L/Z_A$).
Note  from the figure
that there is essentially no dependence on the source for the local
operator.

\if\preprint Y \psoddfigure  486 466 -9 {Figure 9} {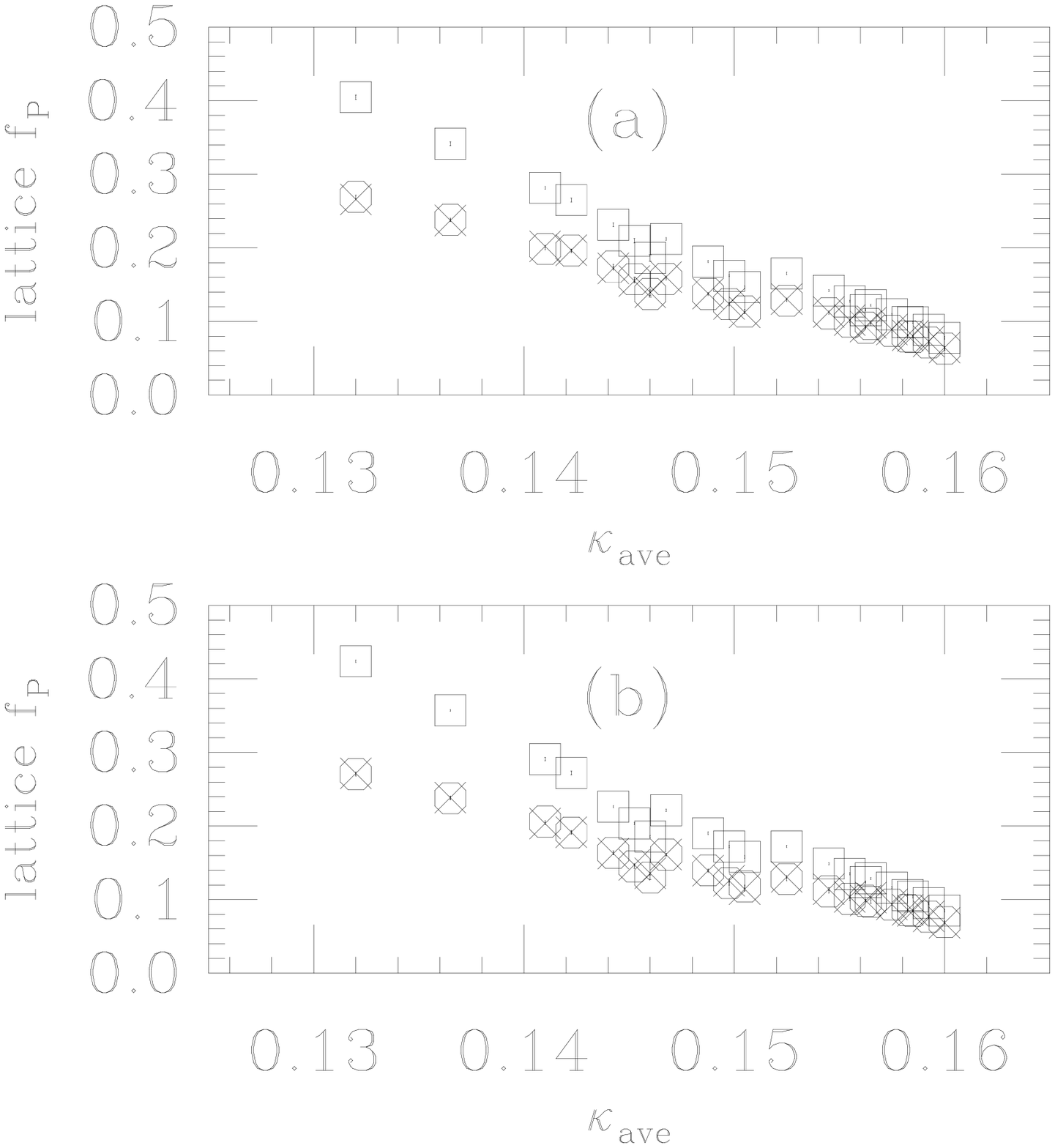} {
Lattice $f_P$ from  axial currents (including all Z-factors from
a tadpole-improved analysis)
as a function of the average hopping parameter, $1/2(\kappa_1 + \kappa_2)$.
Data are labeled with crosses for the local current and a $\gamma_5$
source, octagons for the local current and a $\gamma_0 \gamma_5$ source,
and squares for the nonlocal current.
(a) $am_q=0.01$, (b) $am_q=0.025$.
}\fi

\if\preprint Y \psfigure  252 108 {Figure 10} {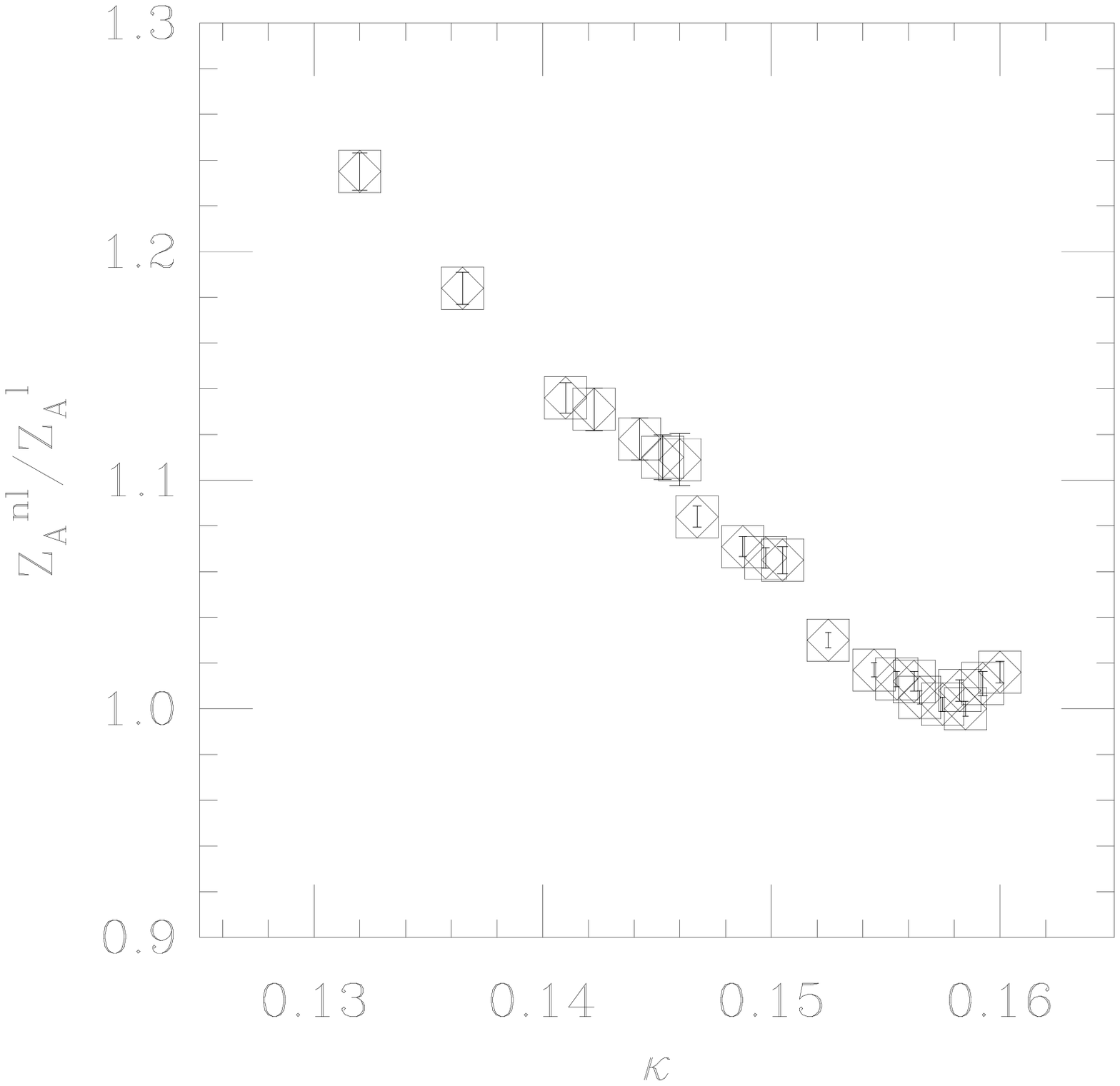} {
Ratios of renormalization factors of nonlocal to local axial
currents.
Results from simulations with sea quark mass $am_q=0.01$ are shown in
squares, and for sea quark mass $am_q=0.025$  in diamonds.
}\fi

We compare the ratio of Z-factors from nonlocal axial to
local axial currents. This ratio is shown in Fig. 10 and Tables XV-XVI.
The expectation of perturbation theory is that this ratio is 1.23,
independent of $\kappa$. The observed $\kappa$ dependence is
larger than for the vector currents, but still only about twenty per cent.
Our data are similar to the quenched $\beta=6$ results of Ref. [\cite{MANDM}],
about 1.25 for the ratio.
Note that the perturbative prediction and the operators themselves are not
``improved.''

Next we wish to compare to experiment. There are two interesting quantities,
$f_\pi$ itself, and the decay constants of heavy pseudoscalar mesons
(such as the D meson). We calculate $f_\pi$ by extrapolating our data for
mesons made of degenerate quarks to $\kappa_c$, using the three lightest
$\kappa$ values. To present a number for $f_D$ we extrapolate light quark-
heavy quark decay constants first to zero light quark mass; then we
 will interpolate in the heavy quark mass.  The extrapolations
are done using jackknifes. We do the extrapolations using both
tadpole improved and conventional renormalizations.

We find four values for $f_\pi$, corresponding to local and nonlocal
 operators with sea quark mass $am_q=0.01$ and 0.025. In lattice units they
are (with a tadpole-improved analysis), and in order local $am_q=.01$,
nonlocal $am_q=.01$,  local $am_q=.025$, nonlocal $am_q=.025$:
 0.053(1), 0.074(1), 0.055(1), and 0.076(1).
With a nominal lattice spacing of $1/a=$ 2 GeV, these numbers are
106(3) MeV, 148(3) MeV, 110(2) MeV, and 153(2) MeV.
The numbers from a conventional analysis are
 0.054(1), 0.089(2), 0.056(1), and 0.092(1),
or
107(3) MeV, 179(3) MeV, 112(2) MeV, and 185(3) MeV.
With our normalization, the experimental number is 132 MeV.
The uncertainties
are entirely from the extrapolation. The lattice spacing is uncertain
to ten or fifteen per cent; this uncertainty dominates a final answer.
We see little variation with sea quark mass; the variation with operator choice
is much greater.

\if\preprint Y \psfigure 252 108  {Figure 11} {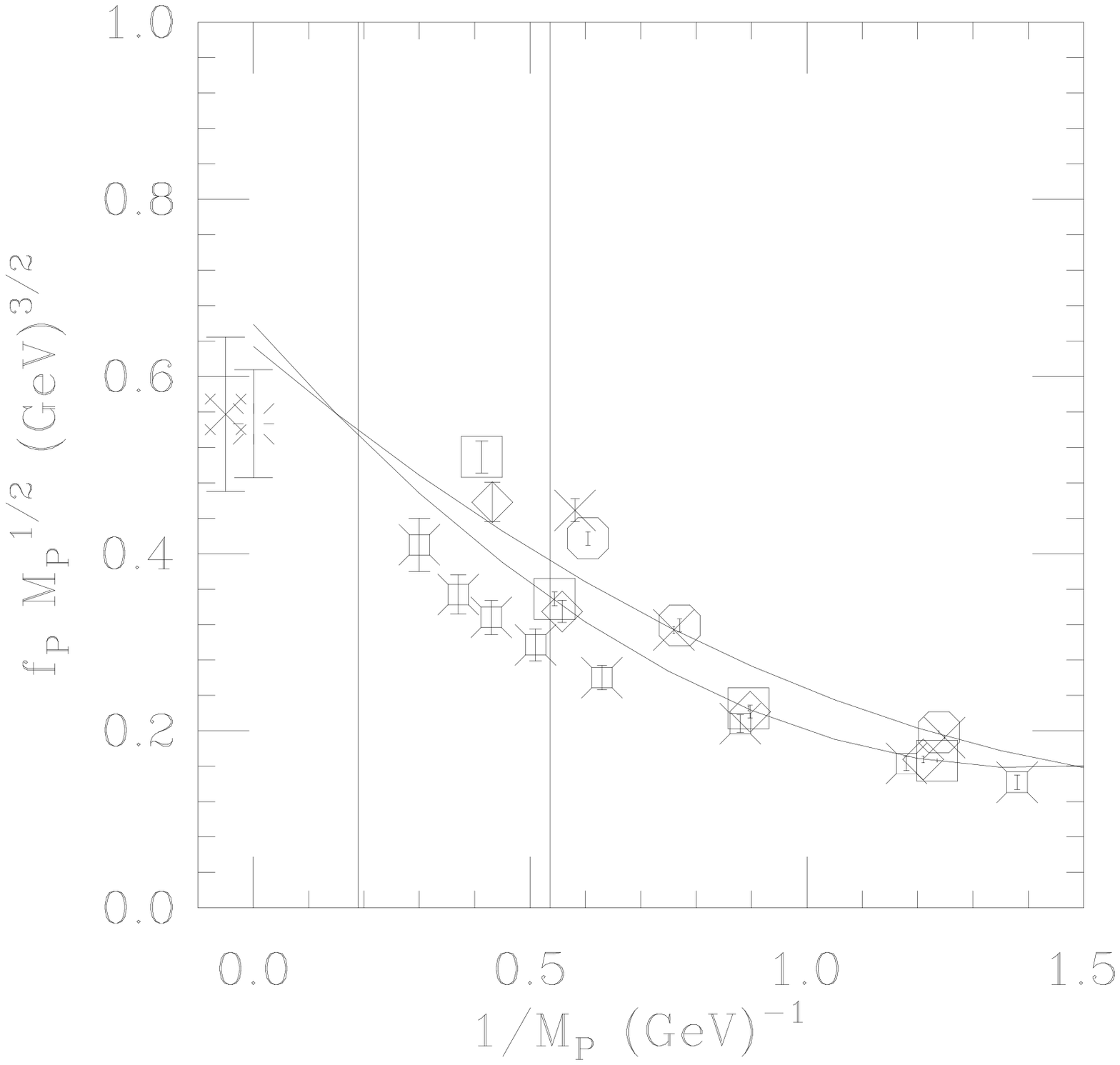}
 {
The quantity $f_P\sqrt{M_P}$ as a function of the inverse pseudoscalar mass,
with lattice data analyzed using tadpole improved perturbation theory.
Data for static quarks are from
Ref. \cite{ALEX} (fancy cross ),
burst is Ref. \cite{ALLTON}.
The fancy squares are the data of Bernard, et. al., Ref. \cite{BLS}
analyzed using ``$\exp(ma)$'' field normalization.
 The scale is set by $f_\pi$.
Our data are  local and nonlocal currents at sea quark mass 0.025
(diamonds and octagons) and local and nonlocal currents at sea quark mass
0.01 (squares and crosses).
The curves are the  quadratic fits described in the text.
The vertical lines identify the points corresponding to $f_B$ and $f_D$.
}\fi

\if\preprint Y \psfigure 252 108  {Figure 12} {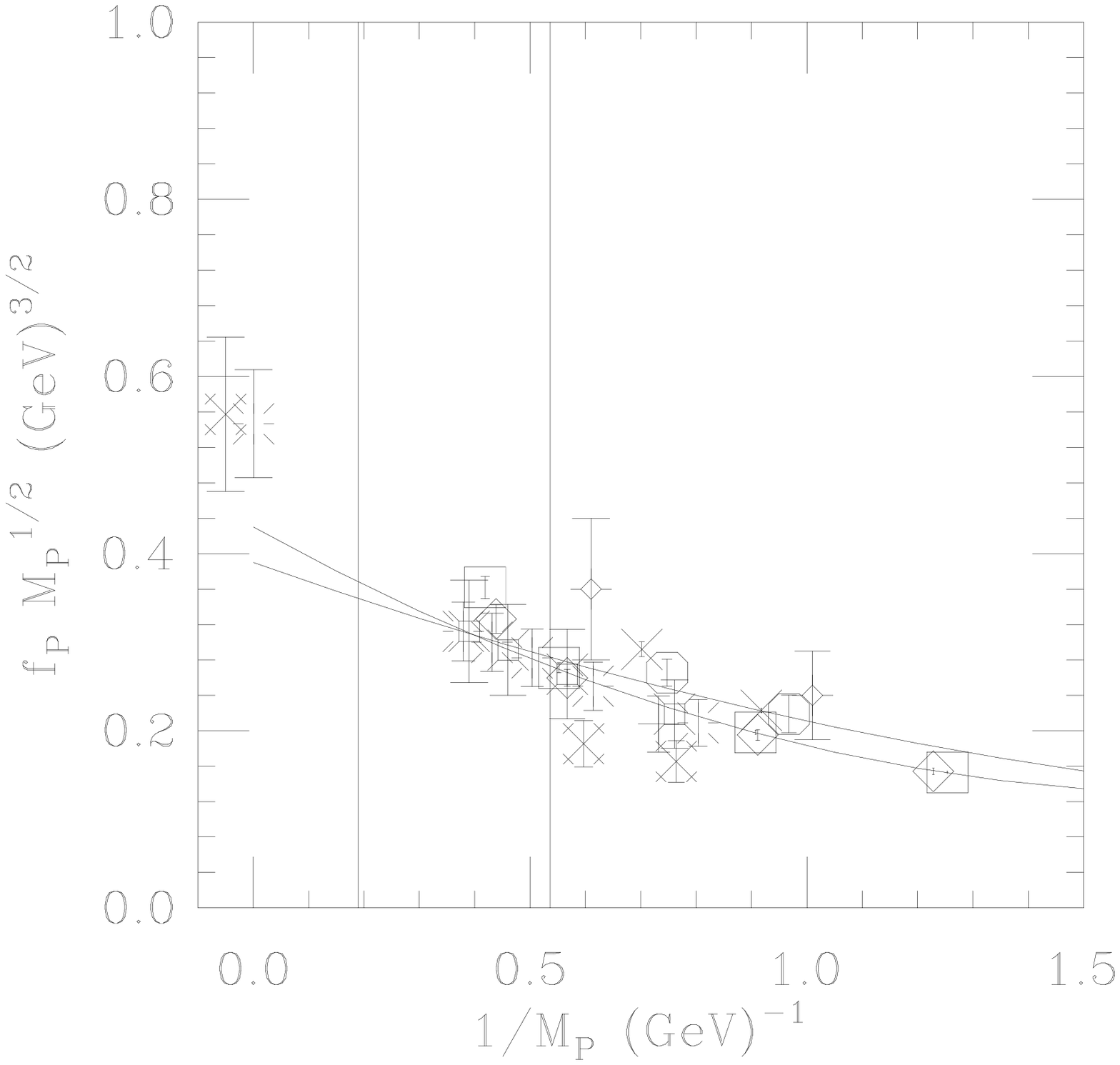}
 {
The quantity $f_P\sqrt{M_P}$ as a function of the inverse pseudoscalar mass,
with lattice data analyzed using conventional field normalization.
Data for static quarks are from
Ref. \cite{ALEX} (fancy cross ),
burst is Ref. \cite{ALLTON}.
Other dynamic heavy quark data are from
the European Lattice Collaboration, Ref. \cite{ELC} (fancy squares),
 Gavela, et. al., Ref. \cite{GAVELA} (plus signs),
and DeGrand and Loft, Ref. \cite{DL} (fancy diamonds)
 The scale is set by $f_\pi$.
Our data are  local and nonlocal currents at sea quark mass 0.025
(diamonds and octagons) and local and nonlocal currents at sea quark mass
0.01 (squares and crosses).
The curves are the  quadratic fits described in the text.
The vertical lines identify the points corresponding to $f_B$ and $f_D$.
}\fi

It has become customary to present heavy-light results as a graph
of $f_P \sqrt{M_P}$ versus the inverse pseudoscalar mass $1/M_P$.
To do this, we need a lattice spacing. We can use either some hadron mass
or the measured $f_\pi$ itself to set the scale. Using $f_\pi$
divides out lattice-to-continuum renormalization factors (at zero quark mass).
Notice with the tadpole improved
 field renormalization the factor $\sqrt{1-6 \tilde \kappa}$
or $2 \kappa \exp(ma)$  is present in the field normalization.
We carry out this extrapolation for each operator at each sea quark mass.
The results of the extrapolations are shown in Tables XVII and
 XVIII and Figs. 11 and 12.
The derived lattice spacings for the nonlocal operators
(with conventional field normalization) are much larger than the nominal
$1/a \simeq 2$ GeV from spectroscopy\rlap.\refto{LATEST},
while the lattice spacings for the conventionally-normalized local operator
and for the tadpole operators are much closer to this number.
This is a reflection of the large $f_\pi$ results for
 this operator and normalization choice reported above.
In the figures we show other
 recent calculations\refto{BLS,GAVELA,ELC,DL}
 of heavy-light decay constants
which as far as we can tell were analyzed similarly to our tadpole
improved or conventional approaches. It is clear from the figures that
our data is rather similar to the results of quenched simulations, when
the analyses are performed in the same way. It is also clear that the
choice of field normalization has a drastic effect on the final answer.
Note from Tables XIII-XIV
 that while we have carried out extrapolations for mesons containing a
$\kappa=0.1320$ quark, the  fits of these quantities before extrapolation have
unacceptably poor confidence levels.

Now we attempt to predict $f_D$. We fit various combinations of our data
points to the form
$$f_P\sqrt{M_P} = A + {B \over M_P} \eqno(7.4)$$
or
$$f_P\sqrt{M_P} = A + {B \over M_P} + {B \over M_P^2}. \eqno(7.5)$$
We discard the heaviest mass points since their confidence level is
poor. A linear extrapolation represents the data poorly.
(None of our heavy quarks are really heavy.) Quadratic
interpolations through all the matrix elements of local operators  or of
all the nonlocal operators
(of both sea quark masses)
are shown in Figs. 11 and 12
(with recent data from Refs. \cite{GAVELA}, \cite{ELC}, \cite{DL},
and \cite{BLS}.
The data of Ref. \cite{BLS} are converted from their figure to
a measurement of $f_P \sqrt{M_P}$ by us\rlap.\refto{HELP}
The conventional analysis undershoots and the tadpole improved analysis
overshoots
the static quark data of Refs. [\cite{ALEX,ALLTON}], which were
not included in the fits.

With the fit, we can then interpolate to the D and B meson mass.
With a tadpole improved analysis
 find $f_D =256(5)$ MeV and $f_B=232(9)$ MeV from the local axial current,
and $f_D =287(4)$ MeV and $f_B=235(5)$ MeV from the nonlocal axial current.
 The uncertainty is completely in the extrapolation.
Interpolating the conventionally normalized lattice data gives
  $f_D =200(4)$ MeV and $f_B=160(7)$ MeV from the local axial current,
and $f_D =208(4)$ MeV and $f_B=152(4)$ MeV from the nonlocal axial current.
We think it is reasonable to include a ten per cent overall uncertainty
just from the lattice spacing in addition to a systematic uncertainty
from the choice of operator. This is about a 20 MeV plus 10 to 20 MeV
effect. The uncertainty due to the choice of $\alpha_s$ is quite small
since it tends to cancel in taking a ratio to $f_\pi$; for
example, in the local axial current with conventional normalization
the variation is 4 MeV when $\alpha_s$ changes from 0.17 to 0.18.

Note that the tadpole-improved data also overshoots the data of Ref.
\cite{BLS}.
This data is from a lattice coupling $\beta=6.3$, or a lattice spacing
$1/a=3.2$ GeV as compared to our $1/a \simeq2$ GeV.
Thus the values of the D meson masses (in lattice units) differ by
about thirty per cent. More importantly, in our simulations the D meson
has a mass in lattice units of about 0.93, which is very heavy. Only for
the lighter masses do our data and that of ref. \cite{BLS} agree;
only for lighter masses do the pseudoscalar masses become small compared
to an inverse lattice spacing.

Recent lattice predictions for $f_D$ (see \cite{BLS,ELC,GAVELA,RECENT})
lie near about 200 MeV. Without more theoretical input we do not know
whether the difference in our predictions with tadpole improvement or with
the conventional normalization is due to a large lattice spacing,
represents a lattice to continuum systematic which should be included in
the overall uncertainty of the lattice prediction, or whether
one method of analysis is to be favored over the other for theoretical
reasons.
Certainly the effects of sea quarks are small compared to other uncertainties.

Finally, we can obtain $f_{D_s}$.
The number has recently been determined by experiment\refto{EXPT}
to be $232 \pm 45 \pm 20 \pm 48$ MeV, and so the tadpole improved numbers
are already in conflict with experiment.
We use
the $\kappa=0.1585$ quark as the strange quark, since the mass of its
vector meson is about 1 GeV (the $\phi$ mass) with an inverse
 lattice spacing of 2 GeV\rlap.\refto{LATEST}
Using the conventional normalization and a lattice spacing determined
by $f_\pi$ we find $f_{D_s}= 220$ MeV from the local operator at
either sea quark mass. We estimate the uncertainty on this number to be
at least
20 MeV from the lattice spacing and 5 MeV in the intrinsic uncertainty
of the lattice measurement.
Decay constants measured with tadpole improvement are much larger
(as they are for $f_D$): for example, the local operator would give 298 MeV
or 264 MeV at $am_q=0.01$ or 0.025; both nonlocal axial current operators
are also quite large, about 320 MeV. This is in conflict with experiment.
Again, we remind the reader that the  masses of these states are
about  1.0 (in lattice units). This is uncomfortably large.

\head{VII. Conclusions}

The data we have presented look rather similar to quenched simulations
at a lattice spacing of $1/a \simeq 2$ GeV. The effects of different mass
sea quarks are small. We see few per cent effects on the vector decay constant,
but any effects of sea quarks in the axial current matrix elements
are masked by uncertainties in the lattice spacing and systematic
differences in the local and nonlocal operators.
The one place we see definite influence of sea quarks is in the
semiperturbative calculation of $\kappa_c$. There a  better  prediction
of $\kappa_c$ needs a coupling constant which evolves under the influence
of two dynamical flavors of quarks. But for experimental observables
the effects of dynamical fermions seem to be subsumed into renormalizations
of lattice parameters. The biggest uncertainties in these calculations,
in fact, do not seem to be the lattice numbers, but the conversion of
lattice numbers into continuum numbers. The only way a pure numerical
simulation
can reduce those systematics is by reducing the lattice spacing to
such a point that the coupling constant is small, independent of
whether sea quarks are present or not.

\vfill\supereject

\head{Acknowledgments}

This work was supported by the U.~S. Department of Energy under contracts
DE--FG02--85ER--40213, 
DE--FG02--91ER--40672, 
DE--AC02--84ER--40125, 
W-31-109-ENG-38, 
and by the National Science Foundation under grants
NSF-PHY87-01775, 
NSF-PHY89-04035  
and
NSF-PHY86-14185. 
The computations were carried out at the
Florida State University Supercomputer Computations
Research Institute which is partially funded by the U.S. Department of
Energy through Contract No. DE-FC05-85ER250000.
 We would like to thank P. Lepage and P. Mackenzie for discussions about
Ref. \cite{PETERPAUL},
J. Fingberg for checking a calculation,
 and J. Labrenz and O. Pene for providing their
data.
We thank  T. Kitchens and J. Mandula for their
continuing support and encouragement.

\if\preprint N
\figurecaptions
\item{1.}
Effective masses for $\kappa=0.1410$ as a function of $r_0$: cross
$r_0=1$, octagon $r_0=2.5$, diamond $r_0=4.0$. Figures are (a) pion,
(b) rho, (c) nucleon, (d) delta.

\item{2.}
Effective masses for $\kappa=0.1585$  as a function of $r_0$: cross
$r_0=3.0$, octagon $r_0=4.0$, diamond $r_0=5.0$.

\item{3.}
Data, and results of a three parameter correlated fit, to
 Gaussian source and sink (crosses), and to a Gaussian source
and  local axial current sink ($\bar \psi \gamma_0 \gamma_5 \psi$) (octagons),
for $am_q=0.01$ sea quark mass and $\kappa=0.1600$ valence quarks.
The second correlator is antiperiodic and its absolute value is shown.

\item{4.}
Results of fits to the lattice matrix element
$\langle 0 | \bar \psi \gamma_0 \gamma_5 \psi | P\rangle$
over the range $t$ to $t_{max}=15$,
from three parameter fits to two propagators.
The diamond, octagon, cross, and square are for heavy quark-light quark
pseudoscalars with heavy $\kappa=$ 0.1320, 0.1410, 0.1525 and 0.1565,
and the burst is for a meson made of two light quarks.
In all cases the light quark hopping parameter is $\kappa=0.1600$.

\item{5.}
Ratios of renormalization factors for (a) local and (b) nonlocal vector
currents to the conserved current.
Results from simulations with sea quark mass $am_q=0.01$ are shown in
squares, and for sea quark mass $am_q=0.025$  in diamonds.

\item{6.}
Lattice $1/f_V$ from the conserved  (Wilson)
vector current, as a function of  $(m_\pi/m_\rho)^2$,
using tadpole
improved perturbation
theory. The labeled points are physical particles
Results from simulations with sea quark mass $am_q=0.01$ are shown in
squares, and for sea quark mass $am_q=0.025$ in diamonds.

\item{7.}
Lattice $1/f_V$ as a function of  $(m_\pi/m_\rho)^2$, using conventional
field normalization and perturbative corrections. The labeled points are
physical particles. Results for the conserved (Wilson) vector current
from simulations with sea quark mass $am_q=0.01$ are shown in
squares, and for sea quark mass $am_q=0.025$ in diamonds.
Results  for the local vector current, scaled by a phenomenological
$Z_V=0.57$
from simulations with sea quark mass $am_q=0.01$ are shown in
fancy crosses, and for sea quark mass $am_q=0.025$ in bursts.
Quenched $\beta=6.0$ from Daniels, et. al., Ref. \cite{GUPTA} are crosses,
and APE results Ref. \cite{APE} are pluses.

\item{8.}
Quark mass in lattice units as a function of   $1/\kappa_{ave} - 1/\kappa_c$.
Results for the local axial current are shown in
squares, and for the nonlocal axial current in diamonds.
The lines show predictions of Eqns. (6.6) and (6.7).
(a) Sea quark mass $am_q=0.01$, (b) sea quark mass $am_q=0.025$.

\item{9.}
Lattice $f_P$ from  axial currents (including all Z-factors from
a tadpole-improved analysis)
as a function of the average hopping parameter, $1/2(\kappa_1 + \kappa_2)$.
Data are labeled with crosses for the local current and a $\gamma_5$
source, octagons for the local current and a $\gamma_0 \gamma_5$ source,
and squares for the nonlocal current.
(a) $am_q=0.01$, (b) $am_q=0.025$.

\item{10.}
Ratios of renormalization factors of nonlocal to local axial
currents.
Results from simulations with sea quark mass $am_q=0.01$ are shown in
squares, and for sea quark mass $am_q=0.025$  in diamonds.

\item{11.}
The quantity $f_P\sqrt{M_P}$ as a function of the inverse pseudoscalar mass,
with lattice data analyzed using tadpole improved perturbation theory.
Data for static quarks are from
Ref. \cite{ALEX} (fancy cross ),
burst is Ref. \cite{ALLTON}.
The fancy squares are the data of Bernard, et. al., Ref. \cite{BLS}
analyzed using ``$\exp(ma)$'' field normalization.
 The scale is set by $f_\pi$.
Our data are  local and nonlocal currents at sea quark mass 0.025
(diamonds and octagons) and local and nonlocal currents at sea quark mass
0.01 (squares and crosses).
The curves are the  quadratic fits described in the text.
The vertical lines identify the points corresponding to $f_B$ and $f_D$.

\item{12.}
The quantity $f_P\sqrt{M_P}$ as a function of the inverse pseudoscalar mass,
with lattice data analyzed using conventional field normalization.
Data for static quarks are from
Ref. \cite{ALEX} (fancy cross ),
squares are Ref. \cite{ALLTON}.
Other dynamic heavy quark data are from
the European Lattice Collaboration, Ref. \cite{ELC} (fancy squares),
 Gavela, et. al., Ref. \cite{GAVELA} (plus signs),
and DeGrand and Loft, Ref. \cite{DL} (fancy diamonds)
 The scale is set by $f_\pi$.
Our data are  local and nonlocal currents at sea quark mass 0.025
(diamonds and octagons) and local and nonlocal currents at sea quark mass
0.01 (squares and crosses).
The curves are the  quadratic fits described in the text.
The vertical lines identify the points corresponding to $f_B$ and $f_D$.

\endfigurecaptions

\head{Table Captions}

\item{I.}
Tadpole-improved renormalization factors for  operators used in
this study.

\item{II.}
Fits to the ratio of conserved vector current to the local vector current
$R^{w,l}$
 with Wilson valence fermions and
$am_q=0.01$  staggered sea quarks.
In this and following tables, numbers in the ``kind'' column
for mesons refers to their quark content: 1 through 6 refer to
hopping parameter 0.1320, 0.1410, 0.1525, 0.1565, 0.1585, and 0.1600.

\item{III.}
Fits to the
 ratio of conserved vector current to the nonlocal vector current $R^{w,l}$
 with Wilson valence fermions and
$am_q=0.01$   staggered sea quarks.

\item{IV.}
Fits to the
 ratio of conserved vector current to the local vector current $R^{w,l}$
 with Wilson valence fermions and
$am_q=0.025$   staggered sea quarks.

\item{V.}
Fits to the
 ratio of conserved vector current to the nonlocal vector current $R^{w,l}$
 with Wilson valence fermions and
$am_q=0.025$   staggered sea quarks.

\item{VI.}
Fits to $f_V$ for the conserved vector current
 with Wilson valence fermions and
$am_q=0.01$  staggered sea quarks.

\item{VII.}
Fits to $f_V$ for the conserved vector current
 with Wilson valence fermions and
$am_q=0.025$  staggered sea quarks.

\item{VIII.}
Fits to the quark mass from the local axial current
 with Wilson valence fermions and
$am_q=0.01$   staggered sea quarks.
 All  Z-factors are set to unity.

\item{IX.}
Fits to the quark mass from the nonlocal axial current
 with Wilson valence fermions and
$am_q=0.01$  staggered sea quarks.
 All  Z-factors are set to unity.

\item{X.}
Fits to the quark mass from the local axial current
 with Wilson valence fermions and
$am_q=0.025$   staggered sea quarks.
 All  Z-factors are set to unity.

\item{XI.}
Fits to the quark mass from the nonlocal axial current
 with Wilson valence fermions and
$am_q=0.025$  staggered sea quarks.
 All  Z-factors are set to unity.

\item{XII.}
Fits to $\kappa_c$ from quark masses calculated from axial current
matrix elements, using pseudoscalar mesons with
all combinations of light valence quarks.

\item{XIII.}
Fits to
pseudoscalar decay constant from local axial currents,
 with Wilson valence fermions and
$am_q=0.01$   staggered sea quarks.
All Z-factors are set to unity.

\item{XIV.}
Fits to
pseudoscalar decay constant from local axial currents,
 with Wilson valence fermions and
$am_q=0.025$  staggered sea quarks.
All Z-factors are set to unity.

\item{XV.}
Fits to
ratio of local to nonlocal axial currents,
 with Wilson valence fermions and
$am_q=0.01$   staggered sea quarks

\item{XVI.}
Fits to
ratio of local to nonlocal axial currents,
 with Wilson valence fermions and
$am_q=0.025$   staggered sea quarks.

\item{XVII.}
Lattice spacing and table of masses and $f_P\sqrt{M_P}$
from axial current matrix elements, from jackknife extrapolations
to zero light quark mass, using tadpole improved perturbation theory.

\item{XVIII.}
Lattice spacing and table of masses and $f_P\sqrt{M_P}$
from axial current matrix elements, from jackknife extrapolations
to zero light quark mass, using conventional $Z$ factors.

\fi

\references

\refis{OVERRELAX}
J. E.~Mandula and M. C.~Ogilvie,  \pl B248, 156, 1990.

\refis{DOUGFIT}
For a discussion of this fitting method see
D. Toussaint, in
``From Actions to Answers--Proceedings of the 1989 Theoretical Advanced
Summer Institute in Particle Physics,'' T. DeGrand and D. Toussaint, eds.,
(World, 1990).

\refis{DEGRANDILU}
T. DeGrand, \journal Comput. Phys. Commun., 52, 161, 1988.

\refis{HEMCGC}
K.~Bitar et al., \prl 65, 2106, 1990, \prd 42, 3794, 1990.

\refis{LIU}
C. Liu, in
 the Proceedings of Lattice '90,
{\sl Nucl. Phys.} {\bf B (Proc. Suppl) 20},  (1991) 149.
  A. D. Kennedy, \journal Intl. J. Mod. Phys., C3, 1, 1992.

\refis{HMD}
H.~C.~Andersen, \journal J. Chem. Phys., 72, 2384, 1980;
S. Duane, \np B257, 652, 1985;
S. Duane and J. Kogut, \prl 55, 2774, 1985;
S. Gottlieb, W. Liu, D. Toussaint, R. Renken and
R. Sugar, \prd 35, 2531, 1987.

\refis{LATEST}
K.~Bitar et al.,  \prd 46, 2169, 1992.

\refis{APE}
S. Cabasino, et. al., \journal Phys. Lett., B258, 195, 1991.
The data for $f_V$  must be corrected by a factor of
$\sqrt{\exp(-m_\rho)}$; private communication to R. Gupta.

\refis{GUPTA}
D. Daniel, R. Gupta, G. Kilcup, A. Patel, and S. Sharpe,
preprint UW-PT-92-05.

\refis{IWASAKI}
Y Iwasaki, K. Kanaya, S. Sakai, and T. Yoshie, \prl 69, 21, 1992.

\refis{DOUGFPI}
C. Bernard, et. al., preprint UCSB-TH-92-30 (1992).

\refis{PETERPAUL}
G. Peter Lepage and Paul B. Mackenzie, FERMILAB preprint FERMILAB-PUB-91/355-T.

\refis{BADHEAVY}
Compare the discussion by S. Hashimoto and Y. Saeki, in
 the Proceedings of Lattice '91, M. Fukugita, Y. Iwasaki, M. Okawa,
and A. Ukawa, eds.
{\sl Nucl. Phys.} {\bf B (Proc. Suppl) 26}, 381 (1992).

\refis{BLS}
C. Bernard, C. Heard, J. Labrenz, and A. Soni, in
 the Proceedings of Lattice '91, M. Fukugita, Y. Iwasaki, M. Okawa,
and A. Ukawa, eds.
{\sl Nucl. Phys.} {\bf B (Proc. Suppl) 26}, 384 (1992),
C. Bernard,  J. Labrenz, and A. Soni, contribution to Lattice 92,
preprint UW/PT-92-21, hep-lat/9211048.

\refis{HELP}
We would like to thank J. Labrenz for discussions about this data.

\refis{MANDM}
L. Maiani and G. Martinelli \journal Phys. Lett., B178, 265, 1986.

\refis{MANDZ}
G. Martinelli and Y-C. Zhang, \journal Phys. Lett., 123B, 433, 1983.

\refis{IMPROVED}
 K. Symanzik, \journal Nucl. Phys., B226, 187, 1983.

\refis{IMPROVEDB}
 G. Martinelli, C. Sachrajda, and
A. Vladikas, in
 the Proceedings of Lattice '90, U. Heller, A. Kennedy and S. Sanielevici, eds.
{\sl Nucl. Phys.} {\bf B (Proc. Suppl) 20},  (1991) 448.

\refis{GROOTHS}
R. Groot, J. Hoek, and J. Smit, \journal Nucl. Phys., B237, 111, 1984.

\refis{URS}
This corrects for the omission of the $n_f$ dependence in [\cite{PETERPAUL}].

\refis{SCALES}
This relation can be inferred from the definition of $\Lambda_V$ in
[\cite{PETERPAUL}] and A.~Billoire, \journal Phys. Lett., 104B, 472, 1981.

\refis{ALEX}
C. Alexandrou, \etal, \journal Phys. Lett., 256B, 60, 1991.

\refis{ALLTON}
C. Allton, \etal, \journal Nucl. Phys., B349, 598, 1991.

\refis{GAVELA}
M. B. Gavela, \etal, \journal Phys. Lett., 206B, 113, 1988.

\refis{ELC}
A. Abada, \etal, in
 the Proceedings of Lattice '91, M. Fukugita, Y. Iwasaki, M. Okawa,
and A. Ukawa, eds.
{\sl Nucl. Phys.} {\bf B (Proc. Suppl) 26}, 344 (1992).

\refis{DL}
T. DeGrand and R, Loft, \journal Phys. Rev., D38, 954, 1988.

\refis{RECENT}
D. G. Roberts, contribution from the UKQCD collaboration reported at
Lattice 92.

\refis{EXPT}
S. Aoki, et.~al., CERN preprint CERN-PPE/92-157 (1992).

\endreferences
\vfill\supereject

\topinsert
\TABLEcap{I}{
Tadpole-improved renormalization factors for  operators used in
this study.
}
$$
{\vbox{\offinterlineskip\halign{
\vrule\fstrut\enskip\hfil#\hfil\enskip&&\fstrut\enskip\hfil#\hfil\enskip\cr
\dbline\notext
operator & name & Z-factor   \endrule
$\bar \psi \gamma_5 \psi$  &  $Z_P $  & $(1 -1.03 \alpha_V){1\over 4}$
\endrule
$\bar \psi \gamma_5\gamma_\mu \psi$&  $Z_A^l $  & $(1 -0.31 \alpha_V){1\over
4}$
\endrule
$\bar \psi \gamma_5\gamma_\mu U_\mu \psi$&  $Z_A^{nl} $  &
$(1 + 0.91 \alpha_V){1\over 4}$
\endrule
$\bar \psi \gamma_\mu \psi$&  $Z_V^{l} $  &
$(1 -0.82 \alpha_V){1\over 4}$
\endrule
$\bar \psi \gamma_\mu U_\mu \psi$&  $Z_V^{nl} $  &
$(1 -1.00 \alpha_V){1\over 4}$
\endrule
$\bar \psi (1+\gamma_\mu) U_\mu \psi$ +\dots &  $Z_V^{w} $  &
${1\over 4}$
\endrule
\notext\sgline\notext
}}}
$$
\vskip 0.2in plus 0.2truein
\endinsert

\topinsert
\TABLEcap{II}{
Fits to the ratio of conserved vector current to the local vector current
$R^{w,l}$
 with Wilson valence fermions and
$am_q=0.01$  staggered sea quarks.
In this and following tables, numbers in the ``kind'' column
for mesons refers to their quark content: 1 through 6 refer to
hopping parameter 0.1320, 0.1410, 0.1525, 0.1565, 0.1585, and 0.1600.
  }
$$
{\vbox{\offinterlineskip\halign{
\vrule\fstrut\enskip\hfil#\hfil\enskip&&\fstrut\enskip\hfil#\hfil\enskip\cr
\dbline\notext
kind & $\kappa_{ave}$ & $D_{min}$ & $D_{max}$ & ratio & $\chi^2$/dof &
 C.L.\endrule
 1 1 & 0.1320 & 11 & 16 & 0.497( 0) &    295.200/9  &      0.000 \endrule
 2 1 & 0.1365 & 11 & 16 & 0.508( 0) &    215.100/9  &      0.000 \endrule
 2 2 & 0.1410 & 7 & 16 & 0.520( 0) &      7.260/17  &      0.950 \endrule
 3 1 & 0.1422 & 11 & 16 & 0.513( 0) &    323.800/9  &      0.000 \endrule
 3 2 & 0.1467 & 7 & 16 & 0.528( 0) &      7.104/17  &      0.955 \endrule
 3 3 & 0.1525 & 8 & 16 & 0.536( 0) &      8.347/15  &      0.820 \endrule
 4 1 & 0.1442 & 11 & 16 & 0.512( 1) &    244.200/9  &      0.000 \endrule
 4 2 & 0.1487 & 7 & 16 & 0.530( 0) &      9.789/17  &      0.833 \endrule
 4 3 & 0.1545 & 8 & 16 & 0.537( 0) &     10.900/15  &      0.619 \endrule
 4 4 & 0.1565 & 11 & 16 & 0.541( 1) &      4.034/9  &      0.776 \endrule
 5 1 & 0.1452 & 11 & 16 & 0.513( 1) &    140.600/9  &      0.000 \endrule
 5 2 & 0.1497 & 8 & 16 & 0.531( 1) &      9.121/15  &      0.764 \endrule
 5 3 & 0.1555 & 11 & 16 & 0.539( 1) &      3.763/9  &      0.807 \endrule
 5 4 & 0.1575 & 11 & 16 & 0.542( 1) &      5.092/9  &      0.649 \endrule
 5 5 & 0.1585 & 11 & 16 & 0.543( 2) &      8.372/9  &      0.301 \endrule
 6 1 & 0.1460 & 11 & 16 & 0.516( 2) &     49.620/9  &      0.000 \endrule
 6 2 & 0.1505 & 8 & 16 & 0.531( 1) &      6.757/15  &      0.914 \endrule
 6 3 & 0.1562 & 10 & 16 & 0.538( 1) &      5.530/11  &      0.786 \endrule
 6 4 & 0.1583 & 11 & 16 & 0.542( 2) &      6.318/9  &      0.503 \endrule
 6 5 & 0.1593 & 11 & 16 & 0.543( 4) &     15.170/9  &      0.034 \endrule
 6 6 & 0.1600 & 8 & 16 & 0.539( 5) &     29.670/15  &      0.005 \endrule
\notext\sgline\notext
}}}
$$
\vskip 0.2in plus 0.2truein
\endinsert

\topinsert
\TABLEcap{III}{
Fits to the
 ratio of conserved vector current to the nonlocal vector current $R^{w,l}$
 with Wilson valence fermions and
$am_q=0.01$   staggered sea quarks.
  }
$$
{\vbox{\offinterlineskip\halign{
\vrule\fstrut\enskip\hfil#\hfil\enskip&&\fstrut\enskip\hfil#\hfil\enskip\cr
\dbline\notext
kind & $\kappa_{ave}$ & $D_{min}$ & $D_{max}$ & ratio & $\chi^2$/dof &
 C.L.\endrule
\ 1 1 & 0.1320 & 11 & 16 & 0.669( 0) &    282.200/7  &      0.000 \endrule
 2 1 & 0.1365 & 5 & 16 & 0.667( 0) &    229.400/19  &      0.000 \endrule
 2 2 & 0.1410 & 8 & 16 & 0.670( 0) &     16.090/13  &      0.244 \endrule
 3 1 & 0.1422 & 11 & 16 & 0.661( 0) &    282.100/7  &      0.000 \endrule
 3 2 & 0.1467 & 11 & 16 & 0.664( 0) &      6.271/7  &      0.508 \endrule
 3 3 & 0.1525 & 11 & 16 & 0.659( 0) &      5.092/7  &      0.649 \endrule
 4 1 & 0.1442 & 5 & 16 & 0.657( 0) &    247.900/19  &      0.000 \endrule
 4 2 & 0.1487 & 11 & 16 & 0.661( 0) &      4.352/7  &      0.738 \endrule
 4 3 & 0.1545 & 10 & 16 & 0.656( 1) &      8.247/9  &      0.509 \endrule
 4 4 & 0.1565 & 10 & 16 & 0.653( 1) &      8.357/9  &      0.499 \endrule
 5 1 & 0.1452 & 9 & 16 & 0.656( 1) &    141.300/11  &      0.000 \endrule
 5 2 & 0.1497 & 11 & 16 & 0.659( 1) &      4.703/7  &      0.696 \endrule
 5 3 & 0.1555 & 10 & 16 & 0.654( 1) &     10.380/9  &      0.321 \endrule
 5 4 & 0.1575 & 10 & 16 & 0.651( 1) &      9.104/9  &      0.428 \endrule
 5 5 & 0.1585 & 10 & 16 & 0.649( 2) &      8.220/9  &      0.512 \endrule
 6 1 & 0.1460 & 10 & 16 & 0.657( 1) &     51.500/9  &      0.000 \endrule
 6 2 & 0.1505 & 11 & 16 & 0.658( 1) &      5.426/7  &      0.608 \endrule
 6 3 & 0.1562 & 10 & 16 & 0.652( 1) &     11.180/9  &      0.264 \endrule
 6 4 & 0.1583 & 10 & 16 & 0.650( 2) &      7.584/9  &      0.577 \endrule
 6 5 & 0.1593 & 8 & 16 & 0.646( 2) &     12.850/13  &      0.459 \endrule
 6 6 & 0.1600 & 6 & 16 & 0.640( 2) &     17.690/17  &      0.409 \endrule
\notext\sgline\notext
}}}
$$
\vskip 0.2in plus 0.2truein
\endinsert

\topinsert
\TABLEcap{IV}{
Fits to the
 ratio of conserved vector current to the local vector current $R^{w,l}$
 with Wilson valence fermions and
$am_q=0.025$   staggered sea quarks.
  }
$$
{\vbox{\offinterlineskip\halign{
\vrule\fstrut\enskip\hfil#\hfil\enskip&&\fstrut\enskip\hfil#\hfil\enskip\cr
\dbline\notext
kind & $\kappa_{ave}$ & $D_{min}$ & $D_{max}$ & ratio & $\chi^2$/dof &
 C.L.\endrule
 1 1 & 0.1320 & 11 & 16 & 0.492( 0) &    303.400/7  &      0.000 \endrule
 2 1 & 0.1365 & 5 & 16 & 0.502( 0) &    256.200/19  &      0.000 \endrule
 2 2 & 0.1410 & 10 & 16 & 0.514( 0) &     10.800/9  &      0.290 \endrule
 3 1 & 0.1422 & 11 & 16 & 0.511( 0) &    307.300/7  &      0.000 \endrule
 3 2 & 0.1467 & 11 & 16 & 0.522( 0) &      7.956/7  &      0.336 \endrule
 3 3 & 0.1525 & 11 & 16 & 0.528( 0) &      5.405/7  &      0.611 \endrule
 4 1 & 0.1442 & 6 & 16 & 0.511( 0) &    248.000/17  &      0.000 \endrule
 4 2 & 0.1487 & 11 & 16 & 0.523( 0) &      5.208/7  &      0.635 \endrule
 4 3 & 0.1545 & 11 & 16 & 0.529( 1) &      6.305/7  &      0.505 \endrule
 4 4 & 0.1565 & 10 & 16 & 0.529( 1) &     11.800/9  &      0.225 \endrule
 5 1 & 0.1452 & 6 & 16 & 0.512( 0) &    148.900/17  &      0.000 \endrule
 5 2 & 0.1497 & 11 & 16 & 0.524( 1) &      4.609/7  &      0.708 \endrule
 5 3 & 0.1555 & 11 & 16 & 0.529( 1) &      9.405/7  &      0.225 \endrule
 5 4 & 0.1575 & 10 & 16 & 0.529( 1) &     12.510/9  &      0.186 \endrule
 5 5 & 0.1585 & 8 & 16 & 0.529( 1) &     16.940/13  &      0.202 \endrule
 6 1 & 0.1460 & 6 & 16 & 0.513( 1) &     60.600/17  &      0.000 \endrule
 6 2 & 0.1505 & 11 & 16 & 0.524( 1) &      4.699/7  &      0.697 \endrule
 6 3 & 0.1562 & 11 & 16 & 0.529( 1) &      9.622/7  &      0.211 \endrule
 6 4 & 0.1583 & 8 & 16 & 0.528( 1) &     16.280/13  &      0.234 \endrule
 6 5 & 0.1593 & 8 & 16 & 0.528( 2) &     13.450/13  &      0.414 \endrule
 6 6 & 0.1600 & 6 & 16 & 0.525( 2) &     19.310/17  &      0.311 \endrule
\notext\sgline\notext
}}}
$$
\vskip 0.2in plus 0.2truein
\endinsert

\topinsert
\TABLEcap{V}{
Fits to the
 ratio of conserved vector current to the nonlocal vector current $R^{w,l}$
 with Wilson valence fermions and
$am_q=0.025$   staggered sea quarks.
  }
$$
{\vbox{\offinterlineskip\halign{
\vrule\fstrut\enskip\hfil#\hfil\enskip&&\fstrut\enskip\hfil#\hfil\enskip\cr
\dbline\notext
kind & $\kappa_{ave}$ & $D_{min}$ & $D_{max}$ & ratio & $\chi^2$/dof &
 C.L.\endrule
 1 1 & 0.1320 & 11 & 16 & 0.675( 0) &    273.400/9  &      0.000 \endrule
 2 1 & 0.1365 & 11 & 16 & 0.673( 0) &    193.900/9  &      0.000 \endrule
 2 2 & 0.1410 & 7 & 16 & 0.675( 0) &      8.975/17  &      0.879 \endrule
 3 1 & 0.1422 & 11 & 16 & 0.664( 0) &    302.400/9  &      0.000 \endrule
 3 2 & 0.1467 & 7 & 16 & 0.670( 0) &      6.298/17  &      0.974 \endrule
 3 3 & 0.1525 & 8 & 16 & 0.666( 0) &      8.032/15  &      0.842 \endrule
 4 1 & 0.1442 & 11 & 16 & 0.657( 1) &    233.300/9  &      0.000 \endrule
 4 2 & 0.1487 & 7 & 16 & 0.667( 0) &      8.239/17  &      0.914 \endrule
 4 3 & 0.1545 & 8 & 16 & 0.664( 0) &     10.480/15  &      0.654 \endrule
 4 4 & 0.1565 & 8 & 16 & 0.663( 1) &     15.910/15  &      0.254 \endrule
 5 1 & 0.1452 & 11 & 16 & 0.655( 1) &    140.100/9  &      0.000 \endrule
 5 2 & 0.1497 & 8 & 16 & 0.665( 1) &      7.415/15  &      0.880 \endrule
 5 3 & 0.1555 & 8 & 16 & 0.663( 1) &     13.730/15  &      0.393 \endrule
 5 4 & 0.1575 & 11 & 16 & 0.663( 2) &      8.265/9  &      0.310 \endrule
 5 5 & 0.1585 & 11 & 16 & 0.662( 2) &     11.480/9  &      0.119 \endrule
 6 1 & 0.1460 & 11 & 16 & 0.655( 3) &     47.740/9  &      0.000 \endrule
 6 2 & 0.1505 & 8 & 16 & 0.664( 1) &      5.736/15  &      0.955 \endrule
 6 3 & 0.1562 & 8 & 16 & 0.662( 1) &     11.280/15  &      0.587 \endrule
 6 4 & 0.1583 & 10 & 16 & 0.660( 3) &     11.440/11  &      0.247 \endrule
 6 5 & 0.1593 & 8 & 16 & 0.657( 3) &     28.220/15  &      0.008 \endrule
 6 6 & 0.1600 & 8 & 16 & 0.656( 6) &     31.820/15  &      0.003 \endrule
\notext\sgline\notext
}}}
$$
\vskip 0.2in plus 0.2truein
\endinsert

\topinsert
\TABLEcap{VI}{
Fits to $f_V$ for the conserved vector current
 with Wilson valence fermions and
$am_q=0.01$  staggered sea quarks.
  }
$$
{\vbox{\offinterlineskip\halign{
\vrule\fstrut\enskip\hfil#\hfil\enskip&&\fstrut\enskip\hfil#\hfil\enskip\cr
\dbline\notext
kind & $\kappa_{ave}$ & $D_{min}$ & $D_{max}$ & $f_V$ & $\chi^2$/dof &
 C.L.\endrule
 1 1 & 0.1320 & 11 & 16 & 0.328( 3) &    180.100/7  &      0.000 \endrule
 2 1 & 0.1365 & 11 & 16 & 0.358( 3) &    117.900/7  &      0.000 \endrule
 2 2 & 0.1410 & 7 & 16 & 0.382( 3) &     13.930/15  &      0.672 \endrule
 3 1 & 0.1422 & 11 & 16 & 0.395( 5) &    198.900/7  &      0.000 \endrule
 3 2 & 0.1467 & 7 & 16 & 0.441( 3) &      8.662/15  &      0.950 \endrule
 3 3 & 0.1525 & 7 & 16 & 0.586( 4) &     12.960/15  &      0.739 \endrule
 4 1 & 0.1442 & 11 & 16 & 0.385( 6) &    162.100/7  &      0.000 \endrule
 4 2 & 0.1487 & 7 & 16 & 0.441( 3) &      7.704/15  &      0.972 \endrule
 4 3 & 0.1545 & 8 & 16 & 0.619( 7) &      7.502/13  &      0.942 \endrule
 4 4 & 0.1565 & 8 & 16 & 0.693( 8) &      9.074/13  &      0.874 \endrule
 5 1 & 0.1452 & 11 & 16 & 0.363( 7) &     95.980/7  &      0.000 \endrule
 5 2 & 0.1497 & 7 & 16 & 0.429( 4) &      8.382/15  &      0.958 \endrule
 5 3 & 0.1555 & 8 & 16 & 0.626( 7) &      8.065/13  &      0.921 \endrule
 5 4 & 0.1575 & 8 & 16 & 0.719( 9) &     11.320/13  &      0.730 \endrule
 5 5 & 0.1585 & 8 & 16 & 0.759(10) &     14.450/13  &      0.492 \endrule
 6 1 & 0.1460 & 7 & 16 & 0.320( 4) &     50.520/15  &      0.000 \endrule
 6 2 & 0.1505 & 7 & 16 & 0.413( 5) &     10.290/15  &      0.891 \endrule
 6 3 & 0.1562 & 8 & 16 & 0.615( 9) &      8.379/13  &      0.908 \endrule
 6 4 & 0.1583 & 8 & 16 & 0.723(11) &     14.060/13  &      0.521 \endrule
 6 5 & 0.1593 & 8 & 16 & 0.774(13) &     20.420/13  &      0.156 \endrule
 6 6 & 0.1600 & 8 & 16 & 0.795(17) &     27.450/13  &      0.025 \endrule
\notext\sgline\notext
}}}
$$
\vskip 0.2in plus 0.2truein
\endinsert

\topinsert
\TABLEcap{VII}{
Fits to $f_V$ for the conserved vector current
 with Wilson valence fermions and
$am_q=0.025$  staggered sea quarks.
  }
$$
{\vbox{\offinterlineskip\halign{
\vrule\fstrut\enskip\hfil#\hfil\enskip&&\fstrut\enskip\hfil#\hfil\enskip\cr
\dbline\notext
kind & $\kappa_{ave}$ & $D_{min}$ & $D_{max}$ & $f_V$ & $\chi^2$/dof &
 C.L.\endrule
 1 1 & 0.1320 & 11 & 16 & 0.316( 3) &    160.400/7  &      0.000 \endrule
 2 1 & 0.1365 & 11 & 16 & 0.343( 3) &    110.300/7  &      0.000 \endrule
 2 2 & 0.1410 & 10 & 16 & 0.372( 3) &      9.104/9  &      0.612 \endrule
 3 1 & 0.1422 & 11 & 16 & 0.370( 4) &    147.800/7  &      0.000 \endrule
 3 2 & 0.1467 & 11 & 16 & 0.421( 5) &      6.977/7  &      0.640 \endrule
 3 3 & 0.1525 & 11 & 16 & 0.547( 8) &      9.368/7  &      0.404 \endrule
 4 1 & 0.1442 & 11 & 16 & 0.357( 5) &    101.400/7  &      0.000 \endrule
 4 2 & 0.1487 & 11 & 16 & 0.421( 6) &      6.547/7  &      0.684 \endrule
 4 3 & 0.1545 & 11 & 16 & 0.586(10) &     11.230/7  &      0.260 \endrule
 4 4 & 0.1565 & 11 & 16 & 0.651(14) &     14.560/7  &      0.104 \endrule
 5 1 & 0.1452 & 11 & 16 & 0.339( 6) &     62.480/7  &      0.000 \endrule
 5 2 & 0.1497 & 11 & 16 & 0.411( 7) &      8.366/7  &      0.498 \endrule
 5 3 & 0.1555 & 11 & 16 & 0.595(12) &     15.330/7  &      0.082 \endrule
 5 4 & 0.1575 & 11 & 16 & 0.674(19) &     16.550/7  &      0.056 \endrule
 5 5 & 0.1585 & 8 & 16 & 0.741(11) &     26.590/13  &      0.032 \endrule
 6 1 & 0.1460 & 4 & 16 & 0.341( 2) &     51.810/21  &      0.001 \endrule
 6 2 & 0.1505 & 11 & 16 & 0.394(10) &     12.590/7  &      0.182 \endrule
 6 3 & 0.1562 & 11 & 16 & 0.593(17) &     16.650/7  &      0.054 \endrule
 6 4 & 0.1583 & 8 & 16 & 0.716(11) &     24.740/13  &      0.054 \endrule
 6 5 & 0.1593 & 10 & 16 & 0.729(21) &     11.660/9  &      0.390 \endrule
 6 6 & 0.1600 & 10 & 16 & 0.732(28) &      8.907/9  &      0.630 \endrule
\notext\sgline\notext
}}}
$$
\vskip 0.2in plus 0.2truein
\endinsert


\topinsert
\TABLEcap{VIII}{
Fits to the quark mass from the local axial current
 with Wilson valence fermions and
$am_q=0.01$   staggered sea quarks.
 All  Z-factors are set to unity.
  }
$$
{\vbox{\offinterlineskip\halign{
\vrule\fstrut\enskip\hfil#\hfil\enskip&&\fstrut\enskip\hfil#\hfil\enskip\cr
\dbline\notext
kind & $\kappa_{ave}$ & $D_{min}$ & $D_{max}$ & $am_q$  & $\chi^2$/dof &
 C.L.\endrule
 1 1 & 0.1320 & 11 & 16 & 0.4997( 3) &    309.300/7  &      0.000 \endrule
 2 1 & 0.1365 & 11 & 16 & 0.4186( 3) &    480.300/7  &      0.000 \endrule
 2 2 & 0.1410 & 8 & 16 & 0.3308( 2) &      6.212/13  &      0.859 \endrule
 3 1 & 0.1422 & 11 & 16 & 0.3204( 3) &   1037.000/7  &      0.000 \endrule
 3 2 & 0.1467 & 9 & 16 & 0.2300( 2) &      3.365/11  &      0.948 \endrule
 3 3 & 0.1525 & 9 & 16 & 0.1307( 2) &      7.943/11  &      0.540 \endrule
 4 1 & 0.1442 & 11 & 16 & 0.2870( 4) &   1016.000/7  &      0.000 \endrule
 4 2 & 0.1487 & 9 & 16 & 0.1963( 2) &     11.340/11  &      0.253 \endrule
 4 3 & 0.1545 & 9 & 16 & 0.0980( 2) &     10.400/11  &      0.319 \endrule
 4 4 & 0.1565 & 11 & 16 & 0.0656( 3) &      3.258/7  &      0.660 \endrule
 5 1 & 0.1452 & 11 & 16 & 0.2772(11) &    985.600/7  &      0.000 \endrule
 5 2 & 0.1497 & 9 & 16 & 0.1800( 2) &     15.920/11  &      0.069 \endrule
 5 3 & 0.1555 & 11 & 16 & 0.0820( 3) &      2.810/7  &      0.729 \endrule
 5 4 & 0.1575 & 11 & 16 & 0.0502( 3) &      2.111/7  &      0.834 \endrule
 5 5 & 0.1585 & 11 & 16 & 0.0351( 4) &      1.361/7  &      0.929 \endrule
 6 1 & 0.1460 & 11 & 16 & 0.2545( 7) &    257.300/7  &      0.000 \endrule
 6 2 & 0.1505 & 9 & 16 & 0.1680( 3) &     12.320/11  &      0.196 \endrule
 6 3 & 0.1562 & 9 & 16 & 0.0708( 3) &      7.980/11  &      0.536 \endrule
 6 4 & 0.1583 & 11 & 16 & 0.0391( 4) &      1.445/7  &      0.919 \endrule
 6 5 & 0.1593 & 11 & 16 & 0.0242( 4) &      1.205/7  &      0.944 \endrule
 6 6 & 0.1600 & 8 & 16 & 0.0133( 3) &      6.659/13  &      0.826 \endrule
\notext\sgline\notext
}}}
$$
\vskip 0.2in plus 0.2truein
\endinsert

\topinsert
\TABLEcap{IX}{
Fits to the quark mass from the nonlocal axial current
 with Wilson valence fermions and
$am_q=0.01$  staggered sea quarks.
 All  Z-factors are set to unity.
  }
$$
{\vbox{\offinterlineskip\halign{
\vrule\fstrut\enskip\hfil#\hfil\enskip&&\fstrut\enskip\hfil#\hfil\enskip\cr
\dbline\notext
kind & $\kappa_{ave}$ & $D_{min}$ & $D_{max}$ & $am_q$ & $\chi^2$/dof &
 C.L.\endrule
 1 1 & 0.1320 & 10 & 16 & 0.6153(54) &    185.200/9  &      0.000 \endrule
 2 1 & 0.1365 & 10 & 16 & 0.4879(37) &    200.800/9  &      0.000 \endrule
 2 2 & 0.1410 & 8 & 16 & 0.3753(21) &     15.270/13  &      0.170 \endrule
 3 1 & 0.1422 & 11 & 16 & 0.3382(24) &    438.400/7  &      0.000 \endrule
 3 2 & 0.1467 & 8 & 16 & 0.2480(11) &     12.100/13  &      0.356 \endrule
 3 3 & 0.1525 & 7 & 16 & 0.1345( 4) &     22.330/15  &      0.050 \endrule
 4 1 & 0.1442 & 11 & 16 & 0.2965(20) &    412.500/7  &      0.000 \endrule
 4 2 & 0.1487 & 7 & 16 & 0.2088( 8) &     13.290/15  &      0.426 \endrule
 4 3 & 0.1545 & 7 & 16 & 0.0997( 3) &     21.590/15  &      0.062 \endrule
 4 4 & 0.1565 & 7 & 16 & 0.0665( 3) &     23.060/15  &      0.041 \endrule
 5 1 & 0.1452 & 11 & 16 & 0.2800(21) &    295.500/7  &      0.000 \endrule
 5 2 & 0.1497 & 7 & 16 & 0.1900( 7) &     14.750/15  &      0.323 \endrule
 5 3 & 0.1555 & 7 & 16 & 0.0833( 3) &     21.360/15  &      0.066 \endrule
 5 4 & 0.1575 & 11 & 16 & 0.0509( 3) &      7.486/7  &      0.187 \endrule
 5 5 & 0.1585 & 7 & 16 & 0.0356( 3) &     21.070/15  &      0.072 \endrule
 6 1 & 0.1460 & 11 & 16 & 0.2678(29) &    135.300/7  &      0.000 \endrule
 6 2 & 0.1505 & 10 & 16 & 0.1774(12) &      5.410/9  &      0.610 \endrule
 6 3 & 0.1562 & 7 & 16 & 0.0715( 3) &     20.130/15  &      0.092 \endrule
 6 4 & 0.1583 & 9 & 16 & 0.0399( 3) &     13.690/11  &      0.134 \endrule
 6 5 & 0.1593 & 7 & 16 & 0.0245( 3) &     17.200/15  &      0.190 \endrule
 6 6 & 0.1600 & 6 & 16 & 0.0132( 3) &     12.630/17  &      0.631 \endrule
\notext\sgline\notext
}}}
$$
\vskip 0.2in plus 0.2truein
\endinsert

\topinsert
\TABLEcap{X}{
Fits to the quark mass from the local axial current
 with Wilson valence fermions and
$am_q=0.025$   staggered sea quarks.
 All  Z-factors are set to unity.
  }
$$
{\vbox{\offinterlineskip\halign{
\vrule\fstrut\enskip\hfil#\hfil\enskip&&\fstrut\enskip\hfil#\hfil\enskip\cr
\dbline\notext
kind & $\kappa_{ave}$ & $D_{min}$ & $D_{max}$ & $am_q$  & $\chi^2$/dof &
 C.L.\endrule
 1 1 & 0.1320 & 4 & 16 & 0.5005( 2) &    249.500/21  &      0.000 \endrule
 2 1 & 0.1365 & 11 & 16 & 0.4202( 3) &    365.600/7  &      0.000 \endrule
 2 2 & 0.1410 & 8 & 16 & 0.3333( 2) &     16.740/13  &      0.116 \endrule
 3 1 & 0.1422 & 11 & 16 & 0.3233( 4) &    792.400/7  &      0.000 \endrule
 3 2 & 0.1467 & 8 & 16 & 0.2328( 2) &     15.030/13  &      0.181 \endrule
 3 3 & 0.1525 & 8 & 16 & 0.1339( 2) &     25.070/13  &      0.009 \endrule
 4 1 & 0.1442 & 11 & 16 & 0.2910( 4) &    769.100/7  &      0.000 \endrule
 4 2 & 0.1487 & 8 & 16 & 0.1995( 2) &     17.890/13  &      0.084 \endrule
 4 3 & 0.1545 & 8 & 16 & 0.1016( 2) &     21.520/13  &      0.028 \endrule
 4 4 & 0.1565 & 8 & 16 & 0.0699( 2) &     17.080/13  &      0.106 \endrule
 5 1 & 0.1452 & 11 & 16 & 0.2741( 5) &    611.500/7  &      0.000 \endrule
 5 2 & 0.1497 & 7 & 16 & 0.1835( 2) &     24.890/15  &      0.024 \endrule
 5 3 & 0.1555 & 8 & 16 & 0.0860( 2) &     18.960/13  &      0.062 \endrule
 5 4 & 0.1575 & 8 & 16 & 0.0547( 2) &     14.820/13  &      0.191 \endrule
 5 5 & 0.1585 & 8 & 16 & 0.0398( 2) &     13.520/13  &      0.261 \endrule
 6 1 & 0.1460 & 11 & 16 & 0.2599( 6) &    381.600/7  &      0.000 \endrule
 6 2 & 0.1505 & 8 & 16 & 0.1718( 3) &     25.160/13  &      0.009 \endrule
 6 3 & 0.1562 & 8 & 16 & 0.0747( 2) &     17.310/13  &      0.099 \endrule
 6 4 & 0.1583 & 8 & 16 & 0.0437( 2) &     13.730/13  &      0.248 \endrule
 6 5 & 0.1593 & 7 & 16 & 0.0291( 2) &     16.060/15  &      0.246 \endrule
 6 6 & 0.1600 & 7 & 16 & 0.0184( 3) &     13.910/15  &      0.380 \endrule
\notext\sgline\notext
}}}
$$
\vskip 0.2in plus 0.2truein
\endinsert

\topinsert
\TABLEcap{XI}{
Fits to the quark mass from the nonlocal axial current
 with Wilson valence fermions and
$am_q=0.025$  staggered sea quarks.
 All  Z-factors are set to unity.
  }
$$
{\vbox{\offinterlineskip\halign{
\vrule\fstrut\enskip\hfil#\hfil\enskip&&\fstrut\enskip\hfil#\hfil\enskip\cr
\dbline\notext
kind & $\kappa_{ave}$ & $D_{min}$ & $D_{max}$ & $am_q$ & $\chi^2$/dof &
 C.L.\endrule
 1 1 & 0.1320 & 6 & 16 & 0.6457(44) &    171.900/17  &      0.000 \endrule
 2 1 & 0.1365 & 6 & 16 & 0.5109(32) &    189.200/17  &      0.000 \endrule
 2 2 & 0.1410 & 9 & 16 & 0.3872(26) &     11.390/11  &      0.250 \endrule
 3 1 & 0.1422 & 11 & 16 & 0.3591(30) &    362.600/7  &      0.000 \endrule
 3 2 & 0.1467 & 8 & 16 & 0.2560(13) &     11.360/13  &      0.414 \endrule
 3 3 & 0.1525 & 9 & 16 & 0.1400( 5) &     20.970/11  &      0.013 \endrule
 4 1 & 0.1442 & 11 & 16 & 0.3148(25) &    364.500/7  &      0.000 \endrule
 4 2 & 0.1487 & 7 & 16 & 0.2163( 9) &     13.000/15  &      0.448 \endrule
 4 3 & 0.1545 & 9 & 16 & 0.1046( 4) &     18.760/11  &      0.027 \endrule
 4 4 & 0.1565 & 9 & 16 & 0.0709( 3) &     13.150/11  &      0.156 \endrule
 5 1 & 0.1452 & 11 & 16 & 0.2941(24) &    306.300/7  &      0.000 \endrule
 5 2 & 0.1497 & 7 & 16 & 0.1973( 8) &     12.790/15  &      0.464 \endrule
 5 3 & 0.1555 & 9 & 16 & 0.0880( 3) &     16.630/11  &      0.055 \endrule
 5 4 & 0.1575 & 9 & 16 & 0.0553( 3) &     10.260/11  &      0.330 \endrule
 5 5 & 0.1585 & 9 & 16 & 0.0401( 3) &      8.020/11  &      0.532 \endrule
 6 1 & 0.1460 & 9 & 16 & 0.2820(22) &    195.000/11  &      0.000 \endrule
 6 2 & 0.1505 & 7 & 16 & 0.1837( 8) &     17.910/15  &      0.161 \endrule
 6 3 & 0.1562 & 8 & 16 & 0.0765( 3) &     18.200/13  &      0.077 \endrule
 6 4 & 0.1583 & 9 & 16 & 0.0442( 3) &      9.140/11  &      0.424 \endrule
 6 5 & 0.1593 & 9 & 16 & 0.0293( 3) &      7.500/11  &      0.585 \endrule
 6 6 & 0.1600 & 9 & 16 & 0.0184( 3) &      7.346/11  &      0.601 \endrule
\notext\sgline\notext
}}}
$$
\vskip 0.2in plus 0.2truein
\endinsert

\topinsert
\TABLEcap{XII}{
Fits to $\kappa_c$ from quark masses calculated from axial current
matrix elements, using pseudoscalar mesons with
all combinations of light valence quarks.
 }
$$
{\vbox{\offinterlineskip\halign{
\vrule\fstrut\enskip\hfil#\hfil\enskip&&\fstrut\enskip\hfil#\hfil\enskip\cr
\dbline\notext
mass & kind & $\kappa_c$ & A   \endrule
 0.01 & local & 0.16091(2) & 0.431(2) \endrule
 0.01 & nonlocal & 0.16090(2) & 0.528(4)    \endrule
 0.025 & local & 0.16128(3) & 0.422(2)    \endrule
 0.025 & nonlocal & 0.16127(4) & 0.518(4)   \endrule
\notext\sgline\notext
}}}
$$
\vskip 0.2in plus 0.2truein
\endinsert


\topinsert
\TABLEcap{XIII}{
Fits to
pseudoscalar decay constant from local axial currents,
 with Wilson valence fermions and
$am_q=0.01$   staggered sea quarks.
All Z-factors are set to unity.
}
$$
{\vbox{\offinterlineskip\halign{
\vrule\fstrut\enskip\hfil#\hfil\enskip&&\fstrut\enskip\hfil#\hfil\enskip\cr
\dbline\notext
kind & $\kappa_{ave}$ & $D_{min}$ & $D_{max}$ & $f_P/Z_A$ & $\chi^2$/dof &
 C.L.\endrule
 1 1 & 0.1320 & 5 & 16 & 0.729( 4) &     95.440/19  &      0.000 \endrule
 2 1 & 0.1365 & 9 & 16 & 0.690( 5) &     51.340/11  &      0.000 \endrule
 2 2 & 0.1410 & 7 & 16 & 0.623( 4) &      8.756/15  &      0.890 \endrule
 3 1 & 0.1422 & 11 & 16 & 0.622( 7) &     80.430/7  &      0.000 \endrule
 3 2 & 0.1467 & 7 & 16 & 0.538( 4) &      5.709/15  &      0.984 \endrule
 3 3 & 0.1525 & 7 & 16 & 0.476( 4) &      7.733/15  &      0.934 \endrule
 4 1 & 0.1442 & 11 & 16 & 0.564( 7) &     74.330/7  &      0.000 \endrule
 4 2 & 0.1487 & 7 & 16 & 0.478( 4) &      6.101/15  &      0.978 \endrule
 4 3 & 0.1545 & 7 & 16 & 0.427( 4) &      9.271/15  &      0.863 \endrule
 4 4 & 0.1565 & 7 & 16 & 0.384( 3) &     11.060/15  &      0.748 \endrule
 5 1 & 0.1452 & 11 & 16 & 0.516( 9) &     60.020/7  &      0.000 \endrule
 5 2 & 0.1497 & 7 & 16 & 0.437( 4) &      7.849/15  &      0.930 \endrule
 5 3 & 0.1555 & 7 & 16 & 0.392( 3) &      9.891/15  &      0.827 \endrule
 5 4 & 0.1575 & 7 & 16 & 0.353( 3) &     11.060/15  &      0.748 \endrule
 5 5 & 0.1585 & 7 & 16 & 0.325( 4) &     10.950/15  &      0.756 \endrule
 6 1 & 0.1460 & 11 & 16 & 0.472(13) &     39.700/7  &      0.000 \endrule
 6 2 & 0.1505 & 7 & 16 & 0.402( 5) &     13.560/15  &      0.559 \endrule
 6 3 & 0.1562 & 7 & 16 & 0.363( 4) &     12.100/15  &      0.671 \endrule
 6 4 & 0.1583 & 6 & 16 & 0.328( 3) &     12.090/17  &      0.795 \endrule
 6 5 & 0.1593 & 6 & 16 & 0.298( 4) &     11.330/17  &      0.839 \endrule
 6 6 & 0.1600 & 5 & 16 & 0.268( 5) &     17.670/19  &      0.545 \endrule
\notext\sgline\notext
}}}
$$
\vskip 0.2in plus 0.2truein
\endinsert

\topinsert
\TABLEcap{XIV}{
Fits to
pseudoscalar decay constant from local axial currents,
 with Wilson valence fermions and
$am_q=0.025$  staggered sea quarks.
All Z-factors are set to unity.
}
$$
{\vbox{\offinterlineskip\halign{
\vrule\fstrut\enskip\hfil#\hfil\enskip&&\fstrut\enskip\hfil#\hfil\enskip\cr
\dbline\notext
kind & $\kappa_{ave}$ & $D_{min}$ & $D_{max}$ & $f_P/Z_A$ & $\chi^2$/dof &
 C.L.\endrule
 1 1 & 0.1320 & 10 & 16 & 0.743( 7) &     91.510/9  &      0.000 \endrule
 2 1 & 0.1365 & 10 & 16 & 0.692( 6) &     61.730/9  &      0.000 \endrule
 2 2 & 0.1410 & 11 & 16 & 0.624( 6) &      1.845/7  &      0.968 \endrule
 3 1 & 0.1422 & 10 & 16 & 0.603( 6) &    106.300/9  &      0.000 \endrule
 3 2 & 0.1467 & 11 & 16 & 0.538( 6) &      3.518/7  &      0.833 \endrule
 3 3 & 0.1525 & 11 & 16 & 0.475( 6) &      5.414/7  &      0.610 \endrule
 4 1 & 0.1442 & 7 & 16 & 0.530( 5) &    112.700/15  &      0.000 \endrule
 4 2 & 0.1487 & 11 & 16 & 0.480( 6) &      5.282/7  &      0.626 \endrule
 4 3 & 0.1545 & 11 & 16 & 0.428( 7) &      6.533/7  &      0.479 \endrule
 4 4 & 0.1565 & 8 & 16 & 0.399( 4) &     18.180/13  &      0.151 \endrule
 5 1 & 0.1452 & 8 & 16 & 0.488( 5) &     78.900/13  &      0.000 \endrule
 5 2 & 0.1497 & 11 & 16 & 0.442( 7) &      5.954/7  &      0.545 \endrule
 5 3 & 0.1555 & 11 & 16 & 0.396( 7) &      7.100/7  &      0.419 \endrule
 5 4 & 0.1575 & 8 & 16 & 0.370( 4) &     16.790/13  &      0.209 \endrule
 5 5 & 0.1585 & 8 & 16 & 0.341( 4) &     15.110/13  &      0.301 \endrule
 6 1 & 0.1460 & 8 & 16 & 0.456( 6) &     47.400/13  &      0.000 \endrule
 6 2 & 0.1505 & 11 & 16 & 0.407( 9) &      6.791/7  &      0.451 \endrule
 6 3 & 0.1562 & 8 & 16 & 0.384( 4) &     17.670/13  &      0.170 \endrule
 6 4 & 0.1583 & 8 & 16 & 0.344( 4) &     15.250/13  &      0.292 \endrule
 6 5 & 0.1593 & 8 & 16 & 0.315( 4) &     13.000/13  &      0.448 \endrule
 6 6 & 0.1600 & 8 & 16 & 0.284( 5) &     10.140/13  &      0.682 \endrule
\notext\sgline\notext
}}}
$$
\vskip 0.2in plus 0.2truein
\endinsert

\topinsert
\TABLEcap{XV}{
Fits to
ratio of local to nonlocal axial currents,
 with Wilson valence fermions and
$am_q=0.01$   staggered sea quarks
}
$$
{\vbox{\offinterlineskip\halign{
\vrule\fstrut\enskip\hfil#\hfil\enskip&&\fstrut\enskip\hfil#\hfil\enskip\cr
\dbline\notext
kind & $\kappa_{ave}$ & $D_{min}$ & $D_{max}$ & ratio & $\chi^2$/dof &
 C.L.\endrule
 1 1 & 0.1320 & 7 & 16 & 1.235( 8) &     87.330/15  &      0.000 \endrule
 2 1 & 0.1365 & 7 & 16 & 1.184( 7) &     56.780/15  &      0.000 \endrule
 2 2 & 0.1410 & 8 & 16 & 1.136( 6) &     14.490/13  &      0.207 \endrule
 3 1 & 0.1422 & 11 & 16 & 1.131( 9) &     69.180/7  &      0.000 \endrule
 3 2 & 0.1467 & 7 & 16 & 1.084( 4) &     17.830/15  &      0.164 \endrule
 3 3 & 0.1525 & 7 & 16 & 1.030( 3) &     13.250/15  &      0.429 \endrule
 4 1 & 0.1442 & 11 & 16 & 1.118( 9) &     61.910/7  &      0.000 \endrule
 4 2 & 0.1487 & 7 & 16 & 1.071( 4) &     15.610/15  &      0.271 \endrule
 4 3 & 0.1545 & 7 & 16 & 1.017( 3) &     12.340/15  &      0.500 \endrule
 4 4 & 0.1565 & 7 & 16 & 1.005( 2) &     13.560/15  &      0.406 \endrule
 5 1 & 0.1452 & 11 & 16 & 1.110( 9) &     44.840/7  &      0.000 \endrule
 5 2 & 0.1497 & 7 & 16 & 1.066( 4) &     14.060/15  &      0.370 \endrule
 5 3 & 0.1555 & 7 & 16 & 1.013( 3) &     12.830/15  &      0.461 \endrule
 5 4 & 0.1575 & 7 & 16 & 1.002( 3) &     15.070/15  &      0.303 \endrule
 5 5 & 0.1585 & 7 & 16 & 1.000( 3) &     16.880/15  &      0.205 \endrule
 6 1 & 0.1460 & 10 & 16 & 1.109(11) &     30.560/9  &      0.000 \endrule
 6 2 & 0.1505 & 7 & 16 & 1.065( 5) &     13.920/15  &      0.380 \endrule
 6 3 & 0.1562 & 7 & 16 & 1.012( 4) &     15.050/15  &      0.304 \endrule
 6 4 & 0.1583 & 10 & 16 & 1.008( 4) &      6.295/9  &      0.506 \endrule
 6 5 & 0.1593 & 10 & 16 & 1.011( 5) &      7.229/9  &      0.405 \endrule
 6 6 & 0.1600 & 6 & 16 & 1.016( 4) &     19.990/17  &      0.172 \endrule
\notext\sgline\notext
}}}
$$
\vskip 0.2in plus 0.2truein
\endinsert

\topinsert
\TABLEcap{XVI}{
Fits to
ratio of local to nonlocal axial currents,
 with Wilson valence fermions and
$am_q=0.025$   staggered sea quarks.
}
$$
{\vbox{\offinterlineskip\halign{
\vrule\fstrut\enskip\hfil#\hfil\enskip&&\fstrut\enskip\hfil#\hfil\enskip\cr
\dbline\notext
kind & $\kappa_{ave}$ & $D_{min}$ & $D_{max}$ & ratio & $\chi^2$/dof &
 C.L.\endrule
 1 1 & 0.1320 & 6 & 16 & 1.291( 9) &     91.110/17  &      0.000 \endrule
 2 1 & 0.1365 & 4 & 16 & 1.211( 5) &     67.170/21  &      0.000 \endrule
 2 2 & 0.1410 & 9 & 16 & 1.162( 8) &     10.670/11  &      0.299 \endrule
 3 1 & 0.1422 & 11 & 16 & 1.149( 9) &     69.450/7  &      0.000 \endrule
 3 2 & 0.1467 & 8 & 16 & 1.098( 5) &     16.010/13  &      0.141 \endrule
 3 3 & 0.1525 & 7 & 16 & 1.044( 3) &     18.140/15  &      0.152 \endrule
 4 1 & 0.1442 & 11 & 16 & 1.128( 8) &     62.960/7  &      0.000 \endrule
 4 2 & 0.1487 & 7 & 16 & 1.085( 4) &     20.200/15  &      0.090 \endrule
 4 3 & 0.1545 & 7 & 16 & 1.030( 2) &     16.240/15  &      0.236 \endrule
 4 4 & 0.1565 & 7 & 16 & 1.017( 2) &     13.140/15  &      0.437 \endrule
 5 1 & 0.1452 & 7 & 16 & 1.133( 6) &     67.940/15  &      0.000 \endrule
 5 2 & 0.1497 & 7 & 16 & 1.079( 4) &     20.050/15  &      0.094 \endrule
 5 3 & 0.1555 & 7 & 16 & 1.025( 2) &     14.480/15  &      0.341 \endrule
 5 4 & 0.1575 & 7 & 16 & 1.013( 2) &     11.300/15  &      0.586 \endrule
 5 5 & 0.1585 & 7 & 16 & 1.009( 2) &      9.734/15  &      0.716 \endrule
 6 1 & 0.1460 & 4 & 16 & 1.133( 4) &     50.410/21  &      0.000 \endrule
 6 2 & 0.1505 & 7 & 16 & 1.075( 4) &     20.680/15  &      0.080 \endrule
 6 3 & 0.1562 & 7 & 16 & 1.023( 2) &     12.970/15  &      0.450 \endrule
 6 4 & 0.1583 & 7 & 16 & 1.011( 2) &      9.669/15  &      0.721 \endrule
 6 5 & 0.1593 & 7 & 16 & 1.009( 2) &      8.447/15  &      0.813 \endrule
 6 6 & 0.1600 & 4 & 16 & 1.013( 2) &     19.330/21  &      0.436 \endrule
\notext\sgline\notext
}}}
$$
\vskip 0.2in plus 0.2truein
\endinsert

\topinsert
\TABLEcap{XVII}{
Lattice spacing and table of masses and $f_P\sqrt{M_P}$
from axial current matrix elements, from jackknife extrapolations
to zero light quark mass, using tadpole improved perturbation theory.
}
$$
{\vbox{\offinterlineskip\halign{
\vrule\fstrut\enskip\hfil#\hfil\enskip&&\fstrut\enskip\hfil#\hfil\enskip\cr
\dbline\notext
$am_q$ &kind      & $1/a$, GeV & mass, GeV & $f_P\sqrt{m_P}$, GeV${}^{3/2}$
    \endrule
 0.01  & local    & 2.51       &  2.42 & 0.509(18)   \endrule
       &          &            &  1.84 & 0.349(8)   \endrule
       &          &            &  1.12 & 0.225(3)   \endrule
       &          &            &  0.81 & 0.166(2)   \endrule
       & nonlocal &  1.80      &  1.72 & 0.449(1)    \endrule
       &          &            &  1.32 & 0.314(4) \endrule
       &          &            &  0.80 & 0.182(2) \endrule
       &          &            &  0.58 & 0.141(2) \endrule
 0.025 & local    & 2.41       &  2.31 & 0.458(22) \endrule
       &          &            &  1.79 & 0.334(12) \endrule
       &          &            &  1.11 & 0.221(7) \endrule
       &          &            &  0.83 & 0.168(4) \endrule
       & nonlocal & 1.74       &  1.65 & 0.417(7) \endrule
       &          &            &  1.30 & 0.319(7) \endrule
       &          &            &  0.81 & 0.198(2) \endrule
       &          &            &  0.60 & 0.145(2) \endrule
\notext\sgline\notext
}}}
$$
\vskip 0.2in plus 0.2truein
\endinsert

\topinsert
\TABLEcap{XVIII}{
Lattice spacing and table of masses and $f_P\sqrt{M_P}$
from axial current matrix elements, from jackknife extrapolations
to zero light quark mass, using conventional $Z$ factors.
}
$$
{\vbox{\offinterlineskip\halign{
\vrule\fstrut\enskip\hfil#\hfil\enskip&&\fstrut\enskip\hfil#\hfil\enskip\cr
\dbline\notext
$am_q$ &kind      & $1/a$, GeV & mass, GeV & $f_P\sqrt{m_P}$, GeV${}^{3/2}$
    \endrule
 0.01  & local    & 2.48       &  2.39 & 0.361(12)   \endrule
       &          &            &  1.81 & 0.271(6)   \endrule
       &          &            &  1.10 & 0.196(2)   \endrule
       &          &            &  0.80 & 0.153(1)   \endrule
       & nonlocal &  1.49      &  1.42 & 0.292(8)    \endrule
       &          &            &  1.09 & 0.224(2) \endrule
       &          &            &  0.66 & 0.155(2) \endrule
       &          &            &  0.48 & 0.118(2) \endrule
 0.025 & local    & 2.38       &  2.28 & 0.326(16) \endrule
       &          &            &  1.76 & 0.260(9) \endrule
       &          &            &  1.09 & 0.195(6) \endrule
       &          &            &  0.81 & 0.154(3) \endrule
       & nonlocal & 1.43       &  1.34 & 0.266(16) \endrule
       &          &            &  1.03 & 0.219(2) \endrule
       &          &            &  0.65 & 0.156(1) \endrule
       &          &            &  0.48 & 0.119(1) \endrule
\notext\sgline\notext
}}}
$$
\vskip 0.2in plus 0.2truein
\endinsert

\vfill\supereject

\endit